%% file: _Main.tex
\documentclass[sigconf]{acmart}
\usepackage{placeins}
\usepackage{float}
\usepackage{microtype}
\usepackage{enumitem}
\usepackage{booktabs}
\usepackage{array}
\usepackage{multirow}
\usepackage{tabularx}
\usepackage{subcaption}
\usepackage{balance}
\usepackage{hyperref}
\usepackage{pgfplots}
\usetikzlibrary{patterns,patterns.meta}
\pgfplotsset{compat=1.18}
\usepackage[normalem]{ulem}
\usepackage{pgfplotstable}
\usepackage{mathtools}
\useunder{\uline}{\ul}{}
\newcolumntype{C}{>{\centering\arraybackslash}p{2cm}}
\hypersetup{
    colorlinks=true,
    linkcolor=blue,
    urlcolor=cyan
}

\usepackage{tikz}
\usepackage{pgfplots}
\pgfplotsset{compat=1.18}



\definecolor{darkblue}{rgb}{0.0, 0.0, 0.55}
\definecolor{ao(english)}{rgb}{0.0, 0.5, 0.0}
\definecolor{coolblack}{rgb}{0.0, 0.18, 0.39}
\definecolor{purpleheart}{rgb}{0.41, 0.21, 0.61}
\definecolor{pastelviolet}{rgb}{0.8, 0.6, 0.79}
\definecolor{lightskyblue}{rgb}{0.53, 0.81, 0.98}
\definecolor{palecornflowerblue}{rgb}{0.67, 0.8, 0.94}
\definecolor{lightmauve}{rgb}{0.86, 0.82, 1.0}
\definecolor{lightpastelpurple}{rgb}{0.69, 0.61, 0.85}
\definecolor{bronze}{rgb}{0.8, 0.5, 0.2}
\definecolor{armygreen}{rgb}{0.29, 0.33, 0.13}
\definecolor{darkpowderblue}{rgb}{0.0, 0.2, 0.6}
\definecolor{falured}{rgb}{0.5, 0.09, 0.09}
\definecolor{outerspace}{rgb}{0.25, 0.29, 0.3}
\definecolor{tangerine}{rgb}{0.95, 0.52, 0.0}
\definecolor{seagreen}{rgb}{0.18, 0.55, 0.34}
\definecolor{springgreen}{rgb}{0.0, 1.0, 0.5}
\definecolor{applegreen}{rgb}{0.55,0.71,0.0}
\definecolor{amethyst}{rgb}{0.6,0.4,0.8}
\definecolor{amber}{rgb}{1.0,0.49,0.0}
\definecolor{darkgreen}{rgb}{0,0.4,0} 

\usepackage[table]{xcolor}
\usepackage{pgfplots}
\usetikzlibrary{plotmarks}
\usepgfplotslibrary{groupplots}

\AtBeginDocument{%
  }

\copyrightyear{2026}
\acmYear{2026}
\setcopyright{cc}
\setcctype{by}
\acmConference[KDD '26]{Proceedings of the 32nd ACM SIGKDD Conference on Knowledge Discovery and Data Mining V.2}{August 09--13, 2026}{Jeju Island, Republic of Korea}
\acmBooktitle{Proceedings of the 32nd ACM SIGKDD Conference on Knowledge Discovery and Data Mining V.2 (KDD '26), August 09--13, 2026, Jeju Island, Republic of Korea}
\acmDOI{10.1145/3770855.3818138}
\acmISBN{979-8-4007-2259-2/2026/08}

\begin{document}

\title{Multi-TAP: Multi-criteria Target Adaptive Persona Modeling \\ for Cross-domain Recommendation}


\author{Daehee Kang}
\affiliation{%
  \institution{Ulsan National Institute of Science \\ and Technology (UNIST)}
  \city{Ulsan}
  \country{Republic of Korea}
}
\email{kangdaehee@unist.ac.kr}

\author{Yeon-Chang Lee}
\authornote{Corresponding author.}
\affiliation{%
  \institution{Ulsan National Institute of Science \\ and Technology (UNIST)}
  \city{Ulsan}
  \country{Republic of Korea}
}
\email{yeonchang@unist.ac.kr}


\begin{abstract}
Cross-domain recommendation (CDR) aims to alleviate data sparsity by transferring knowledge across domains, yet existing methods primarily rely on coarse-grained behavioral signals and often overlook \textit{intra-domain heterogeneity} in user preferences.
We propose \ours, a multi-criteria target-adaptive persona framework that explicitly captures such heterogeneity through semantic persona modeling.
To enable effective transfer, \ours\ selectively incorporates source-domain signals conditioned on the target domain, preserving relevance during knowledge transfer.
Experiments on real-world datasets demonstrate that \ours\ consistently outperforms state-of-the-art CDR methods, highlighting the importance of modeling intra-domain heterogeneity for robust cross-domain recommendation. The codebase of \ours\ is currently available at \url{https://github.com/archivehee/Multi-TAP}.
\end{abstract}



\input{macros}


\maketitle
\section{Introduction} \label{sec:intro} \input{Sections/Intro}
 
\section{Motivation} \label{sec:motiv} \input{Sections/motiv}

\section{The Proposed Framework: \ours} \label{sec:method} \input{Sections/Method}
\section{Evaluation} \label{sec:eval} \input{Sections/Eval}
\section{Related Work} \label{sec:rworks} \input{Sections/RWorks}
\section{Conclusion} \label{sec:concl} \input{Sections/Conclusion}

\begin{acks}
This work was supported by the Institute of Information \& Communications Technology Planning \& Evaluation (IITP) grants funded by the Korea government (MSIT) under the AI Star Fellowship Program (Ulsan National Institute of Science and Technology) (RS-2025-25442824), the Artificial Intelligence Graduate School Program (UNIST) (No. RS-2020-II201336), the Leading Generative AI Human Resources Development Program (IITP-2026-RS-2024-00360227), and the Innovative Human Resource Development for Local Intellectualization Program (IITP-2026-RS-2022-00156361). This work was also supported by the National Research Foundation of Korea (NRF) grant funded by the Korea government (MSIT) (RS-2025-24533171).
\end{acks}
\newpage
\begingroup
\footnotesize
\linespread{0.9}
\selectfont
\bibliographystyle{ACM-Reference-Format}
\bibliography{bibliography}
\endgroup

\appendix \label{sec:app} \input{Sections/Appendix}
\setcounter{table}{0}
\setcounter{figure}{0}


\end{document}

%% file: macros.tex
\newcommand{\spec}{{\it spec.}}
\newcommand{\aka}{{\it a.k.a.}}
\newcommand{\ie}{{\it i.e.}}
\newcommand{\eg}{{\it e.g.}}
\newcommand{\ours}{\textsc{\textsf{Multi-TAP}}}
\newcommand{\blue}{\textcolor{blue}}

\newcommand{\mj}[1]{\textcolor{blue}{[MJ: #1]}}
\newcommand{\jw}[1]{\textcolor{green}{[JW: #1]}}
\newcommand{\yc}[1]{\textcolor{red}{[YC: #1]}}

\newcommand{\oursall}{\textsc{\textsf{{TraceRec(all)}}}}
\newcommand{\oursforward}{\textsc{\textsf{{TraceRec(forward)}}}}
\newcommand{\oursrecent}{\textsc{\textsf{{TraceRec(recent)}}}}

\newcommand{\ourswoproj}{\textsc{\textsf{{TraceRec(w/o proj)}}}}

%% file: Sections/Intro.tex
\noindent\textbf{Background.} 
Cross-domain recommendation (CDR) has emerged as a promising paradigm for alleviating data sparsity in recommendation systems by leveraging user interactions from data-rich source domains to improve recommendation accuracy in data-scarce target domains \cite{CMF, MCF, li2009can, SvyCDR1, EMCDR, zhu2021cross}.
As modern e-commerce platforms (\eg, Amazon, eBay) typically span multiple domains such as electronics, clothing, and books, CDR has been extensively studied as a practical solution beyond single-domain recommendation. 
Most existing CDR approaches \cite{TMCDR, BiTGCF, PTUPCDR} are built upon a \textit{common assumption} that user preferences exhibit a \textit{certain degree of consistency across domains}, such that knowledge learned from a source domain can be effectively transferred to a target domain.


\begin{figure}[t]
\centering
\includegraphics[width=1\linewidth]{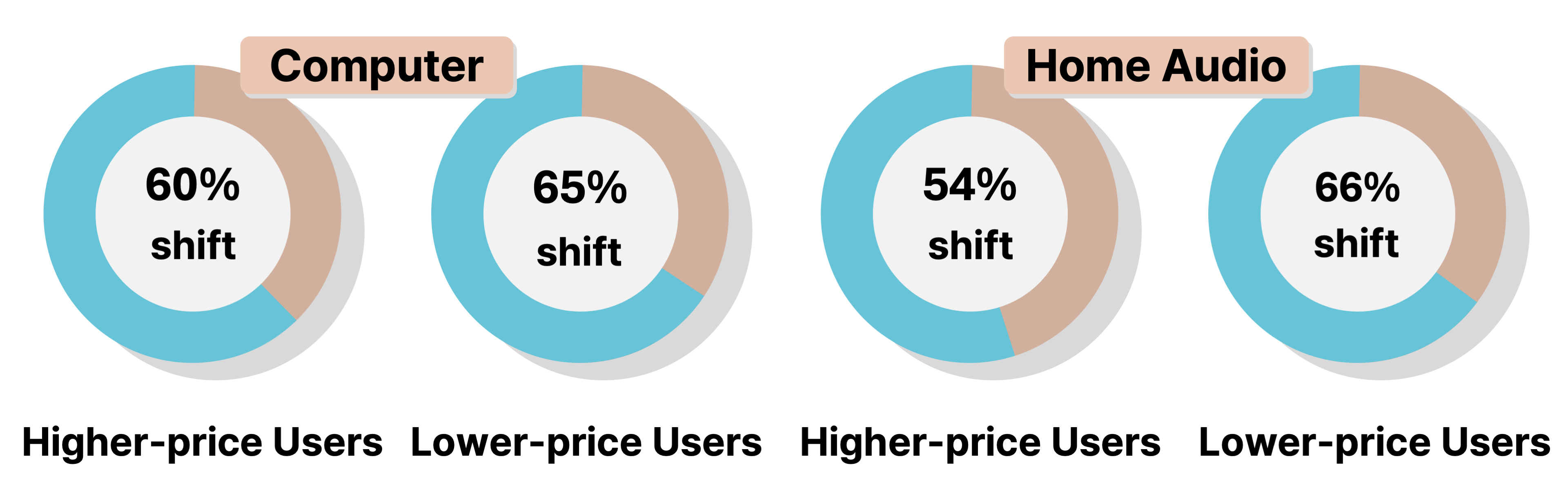}
\vspace{-0.7cm}
\caption{Preference shift ratios across price-based user groups within the Electronics domain.
It reports the proportion of users whose relative price preference shifts when moving from a source category (\textit{Computer} or \textit{Home Audio}) to different target categories, averaged across target categories, using price as an example of a preference criterion.  
}
\label{fig:intro}
\vspace{-0.7cm}
\end{figure}

Under this assumption, prior studies have focused on how to bridge the representational gap between domains, where user representations learned from one domain cannot be directly generalized to another.
Early approaches ~\cite{EMCDR, SA-VAE, SSCDR, DTCDR, GA-DTCDR}, including EMCDR~\cite{EMCDR}  and SA-VAE~\cite{SA-VAE}, rely on overlapping users to \textit{learn mapping functions} that align source- and target-domain user representations.
More recent work~\cite{DisenCDR, UniCDR, DPM-CDR, CAT-ART, EDDA, yang2024not}, including UniCDR~\cite{UniCDR} and EDDA~\cite{EDDA}, adopt preference decomposition or shared latent space learning to \textit{disentangle domain-shared and domain-specific preferences}.
Despite these advances, existing CDR methods primarily \textbf{transfer preferences at the domain level}, relying on a unified representation that aggregates behaviors within each domain, which inevitably compresses context-dependent preference signals.


\vspace{1mm}
\noindent\textbf{Motivation.} 
However, we argue that a more fundamental issue underlies these limitations: the assumption of homogeneous user preferences does not hold even \textit{within} a single domain. 
Our empirical analysis on large-scale real-world datasets shows that users exhibit \textbf{intra-domain heterogeneity}, manifested as divergent consumption behaviors across categories within the same domain (see Figure~\ref{fig:intro}),
as detailed in Section~\ref{sec:motiv}. 
For example, in the Electronics domain, many users who interact with higher-priced items in one category (\eg, \textit{Computer}) frequently shift to budget options in another (\eg, \textit{Home Audio}), while also exhibiting category-specific sensitivity to other \textit{preference criteria} beyond price, such as popularity and ratings.
Given that a single domain in large-scale e-commerce platforms such as Amazon encompasses many fine-grained categories, such preference shifts across categories arise systematically.

These observations motivate a rethinking of the CDR problem.
Rather than asking how to transfer user preferences across domains, the critical question is \textit{which preferences} should be transferred, which in turn requires preserving fine-grained, domain-specific consumption intents.
By overlooking intra-domain heterogeneity, existing CDR methods typically fail to capture fine-grained preference structures in both the source and target domains, leading to substantial information loss even before cross-domain transfer is performed.
Consequently, transferring or sharing these compressed representations across domains further exacerbates this loss, resulting in indiscriminate cross-domain transfer.

\vspace{1mm}
\noindent\textbf{Proposed Ideas.} 
To address these challenges, 
we propose a novel CDR framework, named \textbf{\ours} (\textbf{Multi}-criteria \textbf{T}arget-\textbf{A}dapti\allowbreak ve \textbf{P}ersona).
The key idea of \ours\ is to explicitly decompose user preferences along multiple preference criteria within each domain to construct user personas, and to selectively transfer only the personas that are relevant to the target domain.  
\ours\ consists of two main phases: \textbf{(P1) Multi-criteria Persona Modeling} and \textbf{(P2) Target-adaptive Doppelganger Transfer}.

The first phase, \textbf{Multi-criteria Persona Modeling}, focuses on modeling fine-grained user preferences within each model.
Leveraging user-item interactions and item metadata within each domain, we employ a Large Language Model (LLM) to generate \textit{multiple natural language persona descriptions} for each user, each reflecting distinct consumption intents associated with different preference criteria (\eg, price sensitivity, popularity bias, and category familiarity).
These textual personas are then encoded and aggregated via a self-attention mechanism, producing a single \textit{criterion-aware persona embedding} for each user in each domain.

The second phase, \textbf{Target-adaptive Doppelganger Transfer}, performs selective cross-domain transfer at the persona level.
Given each user's domain-specific persona embeddings obtained from the previous phase, we introduce a \textit{doppelganger persona}, a newly created persona embedding in the source domain designed to mirror the corresponding target-domain persona and enable cross-domain transfer.
By transferring only these doppelganger personas rather than entire source-domain personas, we avoid indiscriminate transfer and better preserve target-domain preference structures. 

Once the source and target domains are specified, \ours\ executes the two phases to construct a target-domain persona embedding for each user, selectively enriched with source-domain information via doppelganger personas.
Final recommendations are generated by computing matching scores between the resulting user representations and target-domain items.

\vspace{1mm}
\noindent\textbf{Contributions.} Our contributions are as follows:
\begin{itemize}[leftmargin=*]
    \item \textbf{Key Empirical Observation.}
    We identify the problem of intra-domain heterogeneity in cross-domain recommendation. 
    \item \textbf{Novel CDR Framework.}
    We propose \ours, a new CDR framework that captures intra-domain heterogeneity via multi-criteria persona modeling and enables selective transfer through target-adaptive doppelganger personas.
    \item \textbf{Extensive Experiments.}
    Extensive experiments on real-world datasets demonstrate that \ours\ consistently outperforms five state-of-the-art CDR methods up to 36.3\% in HR@5.  
\end{itemize}


%% file: Sections/motiv.tex
In this section, we formalize the notion of intra-domain heterogeneity, develop quantitative measures to characterize it, and empirically analyze its prevalence in real-world CDR datasets.

\vspace{1mm}
\noindent\textbf{Definition.} 
In CDR, user preferences are commonly assumed to be homogeneous within each domain, such that a single domain-level representation can adequately summarize a user's behavior \cite{TMCDR, BiTGCF, PTUPCDR}. 
However, this assumption conflicts with behavioral science theories such as
Ajzen's Theory of Planned Behavior ~\cite{plannedbehav}, which argue that observed behaviors arise from multiple latent intentions shaped by context, rather than a single, stable preference.

In recommendation systems, such contextual variation naturally arises even within a single domain, as users may express different preference criteria across categories associated with distinct consumption contexts.
We refer to this phenomenon as \textbf{intra-domain heterogeneity (IDH)}.
IDH implies that aggregating interactions into a single domain-level representation leads to representation compression, masking fine-grained preference signals.
This issue is particularly critical in CDR, where such compressed representations are further transferred across domains.

\begin{figure*}[t]
\centering
\includegraphics[width=\textwidth]{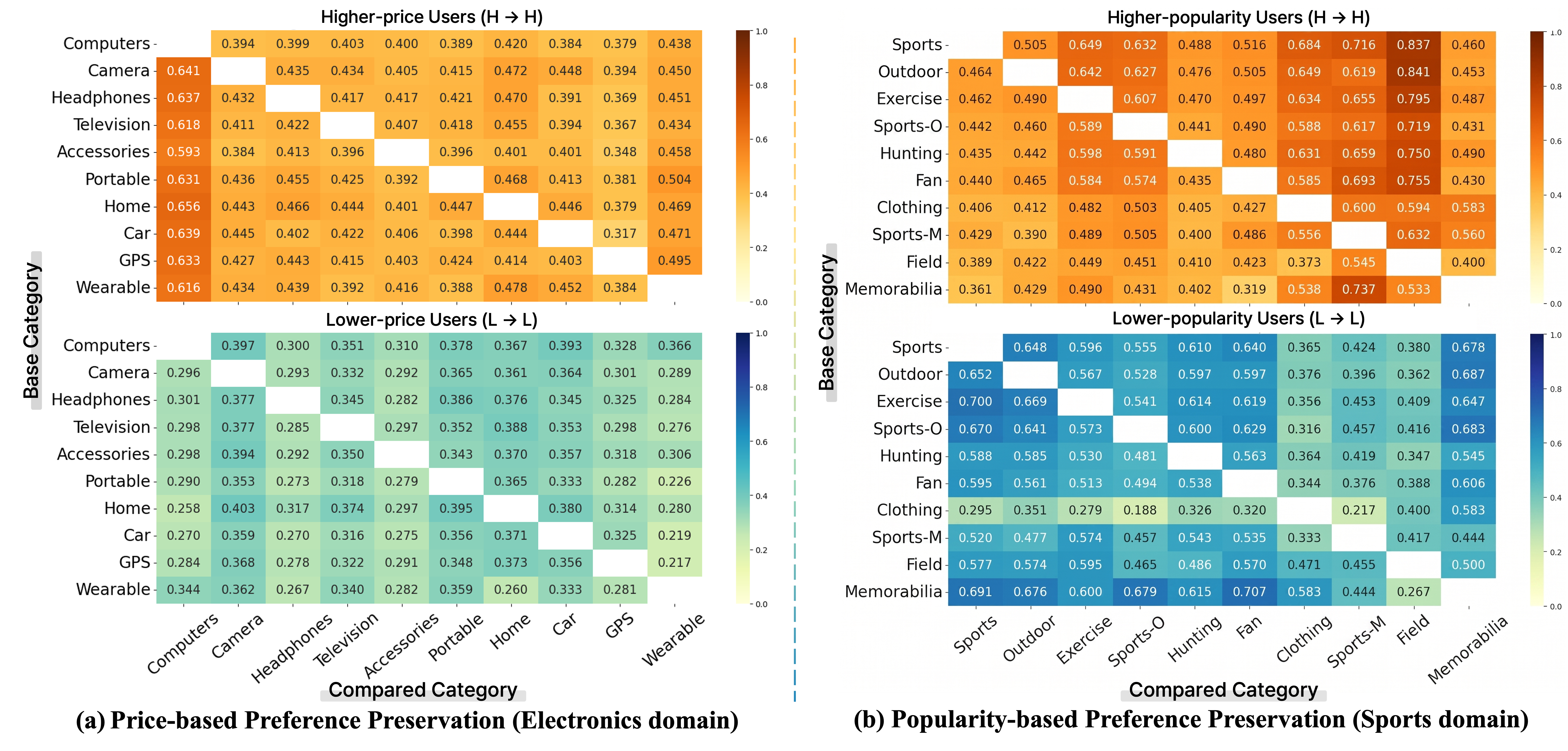}
\vspace{-0.7cm}
\caption{Intra-domain preference heterogeneity across categories.
Each heatmap visualizes the conditional preference preservation ratio, measuring how consistently users maintain their dominant preference label when moving from a base category ($\boldsymbol{y}$-axis) to a compared category ($\boldsymbol{x}$-axis).
Each cell represents the fraction of users who are assigned to a specific preference group (High or Low) in the base category and remain in the same group in the compared category. 
Lower values indicate weaker preference preservation and stronger intra-domain heterogeneity.
}
\label{fig:motiv}
\vspace{-0.4cm}
\end{figure*}

\vspace{1mm}
\noindent\textbf{Measurement.} 
To analyze IDH, we adopt a criterion-specific view of user preferences and develop quantitative measures to characterize category-level preference preservation.

Formally, we consider a domain $d$, associated with a user set $\mathcal{U}_d$, an item set $\mathcal{I}_d$, and a category set $\mathcal{C}_d$, where each item $i \in \mathcal{I}_d$ belongs to a category $c \in \mathcal{C}_d$.
In this section, we focus on a set $\mathcal{K}$ of three \textit{preference criteria}: price sensitivity, quality preference, and popularity bias, each of which is derived from commonly available item-level metadata such as price, average user ratings, and interaction counts, respectively.
For each category $c \in \mathcal{C}_d$ and preference criterion $k \in \mathcal{K}$, we discretize each item $i$'s criterion value into three ordinal bins, denoted as $\text{bin}_i^{(k)}$, based on the category-wise tertiles of the empirical distribution as follows:

\vspace{-0.2cm}
\footnotesize
\begin{equation}\label{eq:bins}
    \text{bin}_i^{(k)} =
    \begin{cases}
    1, & x_i^{(k)} < Q_{c}^{(k)}(1/3), \\
    2, & Q_{c}^{(k)}(1/3) \le x_i^{(k)} < Q_{c}^{(k)}(2/3), \\
    3, & x_i^{(k)} \ge Q_{c}^{(k)}(2/3), 
    \end{cases}
\end{equation}
\normalsize
where $x_i^{(k)}$ denotes the raw metadata value of item $i$ under  $k$, and $Q_{c}^{(k)}(q)$ denotes the $q$-th quantile of the empirical distribution within $c$.
This discretization normalizes criterion scales across categories and enables consistent comparison of preferences.

We then define the \textbf{criterion-specific preference score} $s_{u,c}^{(k)}$ of user $u$ for category $c$ under criterion $k$ as follows: 

\vspace{-0.2cm}
\footnotesize
\begin{equation}
    s_{u,c}^{(k)} = \frac{1}{|\mathcal{I}_{u,c}|} \sum_{i \in \mathcal{I}_{u,c}} \text{bin}_i^{(k)},
\end{equation}
\normalsize
where $\mathcal{I}_{u,c}$ denotes the set of items that user $u$ has interacted with in category $c$, and $\text{bin}_i^{(k)} \in \{1,2,3\}$ denotes the ordinal value of item $i$ under criterion $k$, corresponding to the low, medium, and high bins, respectively.
Intuitively, a larger value of $s_{u,c}^{(k)}$ indicates that user $u$ tends to interact with items associated with higher levels of criterion $k$ within category $c$.
For example, when $k$ corresponds to price, a higher $s_{u,c}^{(k)}$ implies that user $u$ predominantly interacts with relatively high-priced items in category $c$, whereas a lower value indicates a preference for more budget-oriented items.

Based on $s_{u,c}^{(k)}$, we further associate each user $u$ with an \textbf{ordinal preference label} $\ell_{u,c}^{(k)} \in \{\text{Low}, \text{Medium}, \text{High}\}$ for category $c$ under criterion $k$,
where the labels are assigned according to the category-wise tertile thresholds induced by $Q_c^{(k)}(q)$, as used in Eq.~(\ref{eq:bins}).
This label indicates the user's dominant consumption tendency within that category and provides an interpretable abstraction of user preferences at the category level.


Finally, we quantify how consistently users preserve their dominant preference tendencies across categories.
For a criterion $k$ and a label $g \in \{\text{Low}, \text{Medium}, \text{High}\}$, we define the
\textbf{conditional preservation ratio} from category $c_b$ to $c_t$ as follows:

\vspace{-0.3cm}
\footnotesize
\begin{equation}
R_{g}^{(k)}(c_b \rightarrow c_t)
=
\frac{
\sum_{u \in \mathcal{U}_{c_b,c_t}}
\mathbb{I}\!\left(\ell_{u,c_b}^{(k)} = g \ \wedge\  \ell_{u,c_t}^{(k)} = g\right)
}{
\sum_{u \in \mathcal{U}_{c_b,c_t}}
\mathbb{I}\!\left(\ell_{u,c_b}^{(k)} = g\right)
},
\end{equation}
\normalsize
where $\mathcal{U}_{c_b,c_t}$ denotes users who interacted with items in both categories.
The ratio measures the fraction of users labeled as $g$ in $c_b$ who retain the same label in $c_t$ under criterion $k$, and is asymmetric, \ie, $R_{g}^{(k)}(c_b \rightarrow c_t) \neq R_{g}^{(k)}(c_t \rightarrow c_b)$.
Lower values indicate weaker preservation and stronger intra-domain heterogeneity.

\vspace{1mm}
\noindent\textbf{Observations.} 
Figures~\ref{fig:motiv}-(a) and -(b) show price- and popularity-based preference preservation across selected categories in the Electronics and Sports domains, respectively.
For clarity of presentation, we visualize representative category pairs and preference criteria in the main paper, while additional results across other domains, categories, and criteria are provided in \textbf{Appendix~\ref{app_analysis_n_result}}.

As shown in Figure~\ref{fig:motiv}, users’ dominant consumption tendencies are only weakly preserved across categories, revealing substantial \textit{intra-domain heterogeneity}.
Across both domains, preservation ratios are generally low, indicating that user preferences under a given criterion frequently change when moving between categories.
For example, in the Electronics domain, users who exhibit high-price purchasing behavior in one category often fail to maintain the same tendency in others.
In several category pairs, such as \textit{Car} and \textit{GPS}, the preservation ratio drops to around 31\%, suggesting that price sensitivity is highly category-dependent.
This effect is even more pronounced among low-price–oriented users, whose preservation ratios frequently fall below those of high-price users.

These results provide strong empirical evidence of IDH.
Such structural heterogeneity implies that domain-level user representations inevitably compress context-dependent preference signals, underscoring the need for fine-grained, criterion-aware modeling prior to effective cross-domain transfer.

%% file: Sections/Method.tex
\begin{figure*}[t]
\centering
\includegraphics[width=1\linewidth]{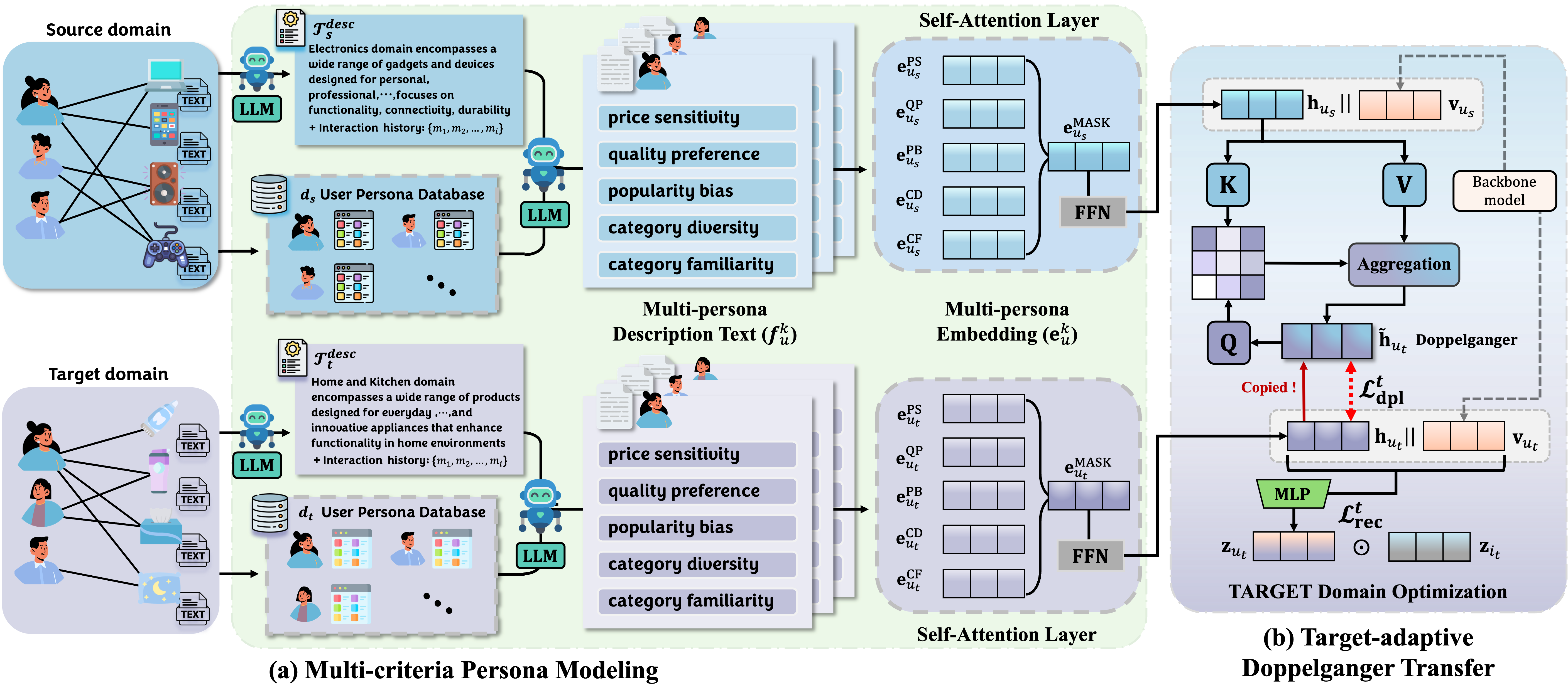}
\vspace{-0.7cm}
\caption{Overview of {\ours}, which consists of (P1) Multi-criteria Persona Modeling, and (P2) Target-adaptive Doppelganger Transfer. 
In (P1), \ours\ decomposes user preferences into explicit, criterion-aware multi-persona embeddings by leveraging a user persona database constructed from interaction histories and item metadata.
In (P2), \ours\ selectively incorporates source-domain information by instantiating doppelganger persona aligned with the target-domain persona embedding.
} 
\label{fig:overview}
\vspace{-0.4cm}
\end{figure*}

In this section, we present {\ours}, a novel framework for cross-domain recommendation.
Section~\ref{sec:method-overview} introduces the overall procedure of \ours, Sections~\ref{sec:method-phase_1} and~\ref{sec:method-phase_2} describe its two core phases, and Section~\ref{sec:method-training} details the training objective.

\subsection{Overview}\label{sec:method-overview} 
We consider a CDR setting with a source domain $d_s$ and a target domain $d_t$, where each domain $d \in \{d_s, d_t\}$ is associated with a user set $\mathcal{U}_d$ and an item set $\mathcal{I}_d$.
Accordingly, we denote the user sets in $d_s$ and $d_t$ as $\mathcal{U}_s$ and $\mathcal{U}_t$, and the corresponding item sets as $\mathcal{I}_s$ and $\mathcal{I}_t$, respectively.
The two domains share a subset of users, denoted as $\mathcal{U}_o = \mathcal{U}_s \cap \mathcal{U}_t$.
Each item $i \in \mathcal{I}_d$ is associated with metadata attributes (\eg, category, price, and store).
Given observed user-item interactions in both domains, the goal of CDR is to improve recommendation accuracy in the target domain $d_t$, particularly for overlapping users $u \in \mathcal{U}_o$, by leveraging information from $d_s$.


Figure~\ref{fig:overview} illustrates the overall architecture of \ours.
The framework consists of two phases: \textbf{(P1) Multi-criteria Persona Modeling} (Figure~\ref{fig:overview}-(a)), which constructs criterion-aware user representations within each domain, and \textbf{(P2) Target-adaptive Doppelganger Transfer} (Figure~\ref{fig:overview}-(b)), which selectively transfers source-domain preference signals in a target-adaptive manner.


In (P1), \ours\ models user preferences within each domain by explicitly decomposing them along multiple preference criteria.
For each overlapping user $u \in \mathcal{U}_o$, we leverage a large language model (LLM) to assist the construction of criterion-specific persona embeddings $\mathbf{e}_{u_s}^{(k)}$ and $\mathbf{e}_{u_t}^{(k)}$ for the source and target domains, along different preference criteria $k \in \mathcal{K}$.
These personas are aggregated into domain-specific persona embeddings $\mathbf{h}_{u_s}$ and $\mathbf{h}_{u_t}$.


In (P2), \ours\ performs selective cross-domain transfer via a target-adaptive doppelganger mechanism.
For each overlapping user $u \in \mathcal{U}_o$, we construct a doppelganger embedding $\mathbf{\tilde{h}}_{u_t}$ in the target domain.
Initialized from $\mathbf{h}_{u_t}$, it selectively absorbs transferable preference signals from the source persona embedding $\mathbf{h}_{u_s}$.
The transfer from $d_s$ to $d_t$ is realized by aligning $\mathbf{h}_{u_t}$ with its doppelganger embedding $\mathbf{\tilde{h}}_{u_t}$ through a contrastive objective.


\input{Table/table-notation}

Finally, \ours\ jointly optimizes the target-domain recommendation and transfer objectives, where recommendations are generated by matching the learned target-domain user representations with item embeddings in the target domain.
Table~\ref{table:notation} summarizes a list of notations used in this paper.


\subsection{Multi-criteria Persona Modeling}\label{sec:method-phase_1} 
The goal of (P1) is to model fine-grained user preferences within each domain by explicitly accounting for intra-domain heterogeneity (IDH).
Rather than representing a user with a single domain-level embedding, we decompose user preferences along multiple interpretable preference criteria and construct criterion-specific personas.
To facilitate this decomposition in a structured and interpretable manner, we leverage an LLM as an auxiliary module for generating preference-aware persona descriptions, rather than as an end-to-end profiling engine.
This phase is performed independently for the source and target domains and provides the foundation for selective cross-domain transfer in the subsequent phase. 
For notational simplicity, we use the notation of a domain $d$, regardless of whether it refers to the source or target domain.


\vspace{1mm}
\noindent\textbf{Step 1: Building User Persona Database.}
We first organize user interaction data into a persona database that serves as the structured input for persona construction.
For each user $u \in \mathcal{U}_d$, we collect the user's interaction history along with the category-level metadata associated with each interaction. 
Based on these metadata, we identify five preference criteria that consistently characterize IDH in real-world e-commerce platforms: \texttt{\small price sensitivity (PS), quality preference (QP), popularity bias (PB), category diversity (CD)}, and \texttt{\small category familiarity (CF)}. 
Since these criteria are derived from commonly available e-commerce metadata (\ie, price, ratings, and item categories), they can be naturally generalized across domains without requiring additional domain-specific features or manual configurations.

Specifically, price, average rating, and interaction counts are observed as category-level metadata and are directly interpreted as user-specific preference tendencies within each category, corresponding to price sensitivity, quality preference, and popularity bias. 
Accordingly, we directly reuse the corresponding ordinal preference labels $\ell_{u,c}^{(k)}$ introduced in Section~\ref{sec:motiv}.

Beyond these item-centric criteria, we further introduce two category-level behavioral criteria, \ie, category diversity and category familiarity, to capture complementary aspects of users' consumption patterns that cannot be inferred from item attributes alone.
We now formalize these category-level behavioral criteria.
\textbf{Category diversity} measures the breadth of a user's consumption by quantifying how many distinct categories the user interacts with within a domain, relative to other users.
Formally, category diversity $x_{u,d}$ for a user $u$ in domain $d$ is defined as follows:

\vspace{-0.2cm}
\footnotesize
\begin{equation}
x_{u,d} = \left| \left\{ \phi(i) \mid i \in \mathcal{I}_{u,d} \right\} \right|,
\end{equation}
\normalsize
where $\phi(i)$ denotes the category of item $i$, and $\mathcal{I}_{u,d}$ denotes the set of items that user $u$ has interacted with in domain $d$.
We then discretize $x_{u,d}$ into three ordinal levels using tertiles of the empirical distribution over all users in domain $d$:

\vspace{-0.2cm}
\footnotesize
\begin{equation}
\ell_{u,d}^\text{CD} =
\begin{cases}
\text{Low}, & x_{u,d} < Q_d(1/3), \\
\text{Medium}, & Q_d(1/3) \le x_{u,d} < Q_d(2/3), \\
\text{High}, &  x_{u,d} \ge Q_d(2/3),
\end{cases}
\end{equation}
\normalsize
where $Q_d(q)$ represents the $q$-th quantile of the empirical distribution of $\{x_{u,d}\}$ over users in domain $d$.
Intuitively, this criterion distinguishes users with narrow category focus from those with broad category exploration within the same domain.

\textbf{Category familiarity} complements category diversity by measuring the depth of a user's interactions with a specific category, relative to other users. 
Since interaction frequencies vary significantly across categories, we normalize users' interaction counts in a category-aware manner.
Specifically, for each category $c$, we define category-specific familiarity thresholds $Q_c(q)$ based on the distribution of user-category interaction counts:

\vspace{-0.2cm}
\footnotesize
\begin{equation}
Q_c(q) = \operatorname{Quantile}_q \left( \left\{ |\mathcal{E}_{u,c}| : u \in \mathcal{U}_c \right\} \right), \quad q \in \left\{\tfrac{1}{3}, \tfrac{2}{3}\right\},
\end{equation}
\normalsize
where $|\mathcal{E}_{u,c}|$ denotes the number of interactions of user $u$ within items in category $c$, and $\mathcal{U}_c$ denotes the set of users who have interacted with items in category $c$.
We then define the category familiarity labels $\ell_{u,c}^\text{CF}$ for user $u$ and category $c$ based on the category-specific tertile thresholds as follows:

\vspace{-0.2cm}
\footnotesize
\begin{equation}
\ell_{u,c}^\text{CF} =
\begin{cases}
\text{Low}, & {|\mathcal{E}_{{u},{c}}}| < Q_c(1/3), \\
\text{Medium}, &  Q_c(1/3) \le {|\mathcal{E}_{{u},{c}}}| < Q_c(2/3), \\
\text{High}, & {|\mathcal{E}_{{u},{c}}}| \ge Q_c(2/3).
\end{cases}
\end{equation}
\normalsize
Intuitively, this criterion captures how deeply a user engages with a category compared to other users consuming the same category.

The user persona database $\mathbb{D}_{u,d}$ for $u$ in $d$ is formally defined as:
\begin{equation}
\mathbb{D}_{u,d} = \left( 
\{\ell_{u,c}^{(k)}\}_{k \in \{\text{PS},\text{QP},\text{PB},\text{CF}\},\, c \in \mathcal{C}_d},\;
\ell_{u,d}^{\text{CD}}\;
\right),
\end{equation}
where $\mathcal{C}_d$ denotes the set of item categories in domain $d$.
That is, $\mathbb{D}_{u,d}$ represents a category-indexed collection of criterion-specific preference signals that explicitly decomposes a user's consumption behavior within domain $d$.


\vspace{1mm}
\noindent\textbf{Step 2: Persona Construction with LLM Assistance.}
In this step, we use an LLM as a controlled auxiliary module to convert structured preference signals into natural-language persona descriptions.
Importantly, the LLM is used \textit{only} to verbalize explicit behavioral statistics, rather than to infer or predict latent user preferences.
Prior studies have shown that long and unstructured inputs (\eg, raw item descriptions or feature lists) can degrade LLM reasoning performance and increase hallucination risks~\cite{levy2024same, huang2025survey, levin2025has, cossio2025comprehensive}.

Motivated by these findings, we adopt a structured prompt design that strictly constrains the LLM's generation space.
Specifically, the LLM input consists of three components:
(1) a concise domain description $\mathcal{T}_{d}^{\text{desc}}$ that specifies domain-specific item semantics,
(2) a small set of representative textual metadata $m_{i,d}$ (\ie, item title and category) associated with items in the user's interaction history, and
(3) the user persona database $\mathbb{D}_{u,d}$, which provides explicit, criterion-specific preference signals at the category level.
In addition, we provide a fixed criterion-aware instruction prompt, denoted as $\mathbb{P}^{\text{sona}}$, which specifies the semantic meaning of each preference criterion and guides the LLM to generate aligned persona descriptions, preventing semantic drift across criteria (see Table~\ref{tab:persona_gen(ours)_prompt_short}).

\input{Table/table-ours_persona_short.tex}

Given this structured input, our objective is not to generate a single holistic user profile, but to obtain \textbf{separate persona descriptions} corresponding to each preference criterion.
Formally, let $\mathcal{K} = \{\text{PS}, \text{QP}, \text{PB}, \text{CD}, \text{CF}\}$ denote the set of preference criteria.
Using a fixed system prompt, the LLM generates all criterion-specific persona descriptions $\{f_u^{(k)}\}_{k \in \mathcal{K}}$ for each user $u$ in a single generation pass, from which each description is obtained as:

\vspace{-0.2cm}
\footnotesize
\begin{equation}
f_{u}^{(k)} = \left[\mathrm{LLM}\!\left(
\mathcal{T}_{d}^{\text{desc}} \,\|\, m_{i,d}\|\,
\left[\,\mathbb{P}^{\text{sona}} \,
\, \|\ \mathbb{D}_{u,d}^{(k)}\right]
\right)\right]_k,
\end{equation}
\normalsize
where $\mathrm{LLM}(\cdot)$ denotes an LLM used for persona generation. 

Importantly, these personas are designed as \textit{distribution-level summaries} of users' accumulated behavioral tendencies under each preference criterion, rather than interaction-level states.
Since such preference distributions typically evolve slowly over time, frequent persona regeneration is unnecessary, enabling efficient and selective updates in practical deployments.

\vspace{1mm}
\noindent\textbf{Step 3: Persona Encoding and Aggregation.}
We encode the criterion-specific persona descriptions into dense vector representations and aggregate them into a unified user embedding.
For each preference criterion $k$, the corresponding persona description is encoded by the LLM text encoder, specifically \texttt{\small text-embedding-3-large}, into a $n$-dimensional embedding $\mathbf{e}_{u}^{(k)} \in \mathbb{R}^n$, yielding a set of decomposed persona embeddings $\mathbf{E}_{u} = \{\mathbf{e}_{u}^{(k)} \mid k \in \mathcal{K}\}$. Using the same LLM family for encoding ensures semantic consistency between the generated persona descriptions and their vector representations.

To aggregate heterogeneous persona signals into a consolidated user representation, we adopt a self-attention-based mechanism \cite{sasrec, zhang2019fdsa, zhao2023alcdr, ju2024drip}.
Specifically, a randomly initialized and learnable mask token $\mathbf{e}_u^{\mathrm{MASK}}$ is used as the aggregation query, while persona embeddings act as keys and values.
For each user $u$ and domain $d$, the attention weight $\alpha_{u}^{(k)}$ for criterion $k$ is computed as follows:


\vspace{-0.2cm}
\footnotesize
\begin{equation}
\alpha_{u}^{(k)} =
\frac{\exp\!\left(({\mathbf{e}_u^{\mathrm{MASK}} \cdot \mathbf{e}_u^{(k)}})/{\sqrt{n}} \right)}
{\sum_{k' \in \mathcal{K}} \exp\!\left( ({\mathbf{e}_u^{\mathrm{MASK}} \cdot \mathbf{e}_u^{(k^{'})}})/{\sqrt{n}} \right)},
\quad 
\end{equation}
\normalsize
where $\cdot$ denotes the dot product and $\sqrt{n}$ is the scaling factor.
Using the resulting attention weights, persona embeddings are aggregated and transformed to obtain the final user embedding $\mathbf{h}_{u}$ as follows:

\vspace{-0.2cm}
\footnotesize
\begin{equation}
\mathbf{h}_{u}
=
\mathrm{Proj}\!\left(
\mathrm{FFN}\!\left(
\sum_{k \in \mathcal{K}} \alpha_{u}^{(k)} \mathbf{e}_{u}^{(k)}
\right)
\right).
\end{equation}
\normalsize
The resulting embedding $\mathbf{h}_{u}$ serves as the corresponding domain persona embedding of user $u$ for downstream recommendation.
Accordingly, we denote the persona-aware embeddings of user $u$ in the source and target domains as $\mathbf{h}_{u_s}$ and $\mathbf{h}_{u_t}$, respectively.

\subsection{Target-adaptive Doppelganger Transfer}\label{sec:method-phase_2}
The goal of (P2) is to perform target-adaptive knowledge transfer by leveraging the persona-aware user representations generated in (P1).
While contrastive learning has been widely used in CDR to align overlapping users across domains \cite{yang2024not, hao2024motif, zhao2023coast, zhou2025ctt}, \textit{direct source-target alignment} can compress domain-specific preferences and distort target-domain representations.
To address this limitation, we design a target-centered transfer mechanism that is trained exclusively on the target domain. 
Specifically, we introduce a doppelganger-based transfer framework, where each target-domain persona embedding serves as the anchor.
The anchor selectively absorbs transferable information from its source-domain counterpart through a target-conditioned intermediate representation, rather than enforcing direct source-target embedding collapse.


\vspace{1mm}
\noindent\textbf{Step 1: Doppelganger Generation.}
We construct a \textbf{doppelganger embedding} for each user $u_t \in \mathcal{U}_t$ as an intermediate representation for cross-domain transfer.
Given the target persona embedding $\mathbf{h}_{u_t}$ obtained in (P1), we initialize the doppelganger embedding $\tilde{\mathbf{h}}_{{u_t}}$ as a direct copy of $\mathbf{h}_{u_t}$, ensuring a target-centered starting point that preserves target-domain preference characteristics.

To selectively incorporate transferable information from the source domain $d_s$, we refine $\tilde{\mathbf{h}}_{{u_t}}$ via a cross-attention operation between the target and source persona embeddings of the same user.
Specifically, the initial target embedding $\tilde{\mathbf{h}}_{{u_t}}$ serves as the query, while the corresponding source persona embedding $\mathbf{h}_{u_s}$ acts as the key and value.
The resulting cross-attention weight $\alpha_u^{\mathrm{dpl}}$ and the refined doppelganger embedding $\tilde{\mathbf{h}}_{u_t}$ are computed as follows:


\vspace{-0.2cm}
\footnotesize
\begin{equation}
\alpha_{u,s}^{\text{dpl}}
=
\frac{
\exp\!\left( (\tilde{\mathbf{h}}_{u_t} \cdot \mathbf{h}_{u_s}) / \sqrt{n} \right)
}{
\sum_{{u}_{s}' \in {\mathcal{U}_s}}
\exp\!\left( (\tilde{\mathbf{h}}_{u_t} \cdot \mathbf{h}_{u_{s}'}) / \sqrt{n} \right)
},
\qquad
\tilde{\mathbf{h}}_{u_t}
=
\alpha_{u,s}^{\text{dpl}} \mathbf{h}_{u_s},
\end{equation}
\normalsize
where $\tilde{\mathbf{h}}_{u_t}$ in the attention score denotes the initial target persona embedding before aggregation.
This design allows the doppelganger persona to selectively absorb source-domain signals in a target.



\vspace{1mm}
\noindent\textbf{Step 2: Doppelganger-based Knowledge Transfer.}
In this step, we perform knowledge transfer by aligning each target persona embedding $\mathbf{h}_{{u_t}}$ with its corresponding doppelganger embedding $\tilde{\mathbf{h}}_{u_t}$. 
Specifically, we encourage $\mathbf{h}_{{u_t}}$ to be close to $\tilde{\mathbf{h}}_{u_t}$, while pushing it away from the doppelgangers $\tilde{\mathbf{h}}_{u'_t}$ of other users using an InfoNCE-based contrastive objective \cite{pmlr-v119-simclr, zhou2025ctt}:

\vspace{-0.2cm}
\footnotesize
\begin{equation}
\mathcal{L}_{\mathrm{dpl}}
=
- \sum_{u \in \mathcal{U}_o}
\log
\frac{
\exp\!\left(
\operatorname{sim}(\mathbf{h}_{u_t}, \tilde{\mathbf{h}}_{u_t}) / \tau
\right)
}{
\sum_{u' \in \mathcal{U}_o}
\exp\!\left(
\operatorname{sim}(\mathbf{h}_{u_t}, \tilde{\mathbf{h}}_{u'_{t}}) / \tau
\right)
},
\end{equation}
\normalsize
where $\operatorname{sim}(\cdot,\cdot)$ denotes cosine similarity and $\tau$ is a temperature hyperparameter.
As a result, transferable source-domain signals are incorporated indirectly, \textit{without enforcing direct source–target embedding alignment}.
Importantly, this contrastive objective is defined on the target domain and optimized jointly with the recommendation loss described in the next section. 






\subsection{Training Process}\label{sec:method-training} 
In recommendation tasks, semantic persona embeddings alone are insufficient to capture collaborative signals from user-item interactions.
We therefore complement persona embeddings with ID-based behavioral embeddings and jointly optimize them under the recommendation objective.
In the following, we describe the training process of \ours\ from the perspective of the target domain, which constitutes the main optimization stage of our framework.

\vspace{1mm}
\noindent\textbf{Target-domain Recommendation Training.}
For the target domain $d_t$, we construct the final representation by combining the target persona embedding with collaborative behavioral signals.
Specifically, we independently pretrain a LightGCN \cite{he2020lightgcn} on the target-domain user-item interaction graph to obtain ID-based user and item embeddings $\mathbf{v}_{u_{t}}$ and $\mathbf{v}_{i_{t}}$, which serve as complementary behavioral backbone features. 
The final user and item representations are then constructed as follows:

\vspace{-0.2cm}
\footnotesize
\begin{equation}
\mathbf{z}_{u_{t}} = \mathrm{MLP}\!\left([\mathbf{h}_{u_{t}} \,\|\, \mathbf{v}_{u_{t}}]\right), \quad
\mathbf{z}_{i_{t}} = \mathrm{MLP}\!\left([\mathbf{h}_{i_{t}} \,\|\, \mathbf{v}_{i_{t}}]\right),
\end{equation}
\normalsize
where $\mathbf{h}_{i_{t}}$ denotes the semantic embedding of item $i_t$, encoded from its textual metadata (\ie, $\mathcal{T}^{desc}_d$ and item category) using the same text encoder as in (P1), 
and $\mathrm{MLP}$ denotes a single-layer feed-forward network.
This design enables \ours\ to incorporate persona semantics into standard ID-based recommendation architectures.

Finally, the recommendation objective on the target domain $d_t$ is defined using the Bayesian Personalized Ranking (BPR) \cite{BPR} loss:

\vspace{-0.2cm}
\begin{equation}\label{eq:BPR}
\mathcal{L}^{t}_{\mathrm{rec}}
= - \sum_{u \in \mathcal{U}_{t}}
\! \sum_{(i,i') \in \mathcal{I}_{u,t}}
\log \sigma\!\left(\hat{y}^{d_t}_{u,i} - \hat{y}^{d_t}_{u,i'}\right)
\ + \lVert \Theta \rVert^2,
\end{equation}
\normalsize
where $\sigma(\cdot)$ denotes the sigmoid function, $\Theta$ represents model parameters subject to weight decay, and
$(i,i')$ denotes a training pair for user $u$, consisting of a positive item $i$ and a sampled negative item $i'$ from the target domain.
The preference score $\hat{y}_{u,i}^{d_t}= \mathbf{z}_{u_{t}} \cdot \mathbf{z}_{i_{t}}$ is computed via inner-product matching.

\input{Table/table-dataset.tex}

\vspace{1mm}
\noindent\textbf{Joint Optimization with Target-adaptive Transfer.}
To enable target-adaptive knowledge transfer, we further incorporate the doppelganger-based contrastive objective introduced in Section~\ref{sec:method-phase_2}.
The overall training objective on $d_t$ is jointly optimized as follows:

\vspace{-0.2cm}
\footnotesize
\begin{equation}
\mathcal{L}_{\mathrm{total}}^{t}
=
\mathcal{L}_{\mathrm{rec}}^{t}
+
\lambda \cdot \mathcal{L}_{\mathrm{dpl}},
\end{equation}
\normalsize
where $\lambda$ controls the contribution of the doppelganger-based InfoNCE loss.
By jointly optimizing the recommendation and contrastive objectives, \ours\ effectively leverages collaborative supervision while selectively integrating transferable source-domain signals in a target-centered manner.

\input{Table/table-eq1.tex}

\vspace{1mm}
\noindent\textbf{Source-domain Training as a Prerequisite.}
Finally, we note that the recommendation objective on the source domain $d_s$ is optimized independently as a prerequisite for knowledge transfer.
Specifically, the same BPR objectives defined in Eq.~(\ref{eq:BPR}) is applied to the source domain, with domain-specific inputs and indices (\ie, $d_t$ replaced by $d_s$).
This step provides source persona embeddings $\mathbf{h}_{u_{s}}$ and $\mathbf{h}_{i_{s}}$ obtained via (P1), and behavioral embeddings $\mathbf{v}_{u_{s}}$ and $\mathbf{v}_{i_{s}}$ obtained via LightGCN.
However, these embeddings do not participate in the joint optimization on the target domain.
In practice, source-domain training can be performed once a domain pair is determined, or reused from a pretrained model when available.

%% file: Table/table-notation.tex
\begin{table}[t]
\centering
\small
\caption{Key notations used in this paper}
\vspace{-0.25cm}
\label{table:notation}

\resizebox{.48\textwidth}{!}{
\renewcommand{\arraystretch}{1.2}
\setlength{\tabcolsep}{6pt}
\begin{tabular}{p{0.2\linewidth}|p{0.95\linewidth}}
\toprule
\hfil\textbf{Notation} & \multicolumn{1}{c}{\textbf{Description}}\\
\midrule

\hfil $d_s, d_t$ & Source and target domains \\
\hfil $\mathcal{U}_s, \mathcal{U}_t, \mathcal{U}_o$ & User sets in the source and target domains, and the overlapping user set \\
\hfil $\mathcal{I}_s, \mathcal{I}_t$ & Item sets in the source and target domains \\
\hfil $\mathcal{C}_s$, $\mathcal{C}_t$ & Set of item categories in source and target domains \\
\hfil $\mathcal{K}$ & Set of preference criteria\\
\midrule
\hfil $\mathbb{D}_{u,s}, \mathbb{D}_{u,t}$ & User persona database in the source and target domains\\
\hfil $\mathbf{e}_{u_s}^{(k)}, \mathbf{e}_{u_t}^{(k)}$ & Criterion-specific user embeddings in the source and target domains \\
\hfil $\mathbf{h}_{u_s}, \mathbf{h}_{u_t}$ & Persona-aware user embeddings in the source and target domains \\
\hfil $\mathbf{h}_{i_s}, \mathbf{h}_{i_t}$ & Semantic item embeddings in the source and target domains \\
\hfil $\mathbf{\tilde{h}}_{u_t}$ & Target-adaptive doppelganger embedding of user $u$ \\




\bottomrule
\end{tabular}
}

\vspace{-0.5cm}
\end{table}

%% file: Table/table-ours_persona_short.tex
\begin{table}[t]
\caption{Structured LLM prompt for multi-persona generation in \ours. The prompt decomposes user preferences into five criterion-specific personas using structured behavioral signals and interaction metadata.}
\label{tab:persona_gen(ours)_prompt_short}
\centering
\scriptsize
\renewcommand{\arraystretch}{1.5}
\begin{tabularx}{\columnwidth}{l X}
\hline
\rowcolor[HTML]{F2F2F2}
\textbf{Component} & \textbf{Content} \\ \hline
\textbf{Task Description} & You are an e-commerce user analysis expert. Analyze the provided context (Domain, User ID, USER SPECIFIC INPUTS, and History) to generate FIVE natural-language personas, each focusing on: \newline
1. Price sensitivity, 2. Quality preference, 3. Popularity bias, 4. Category familiarity, 5. Category diversity. \\ \hline

\textbf{Input Format} & 
\texttt{\{ "Domain": "<domain\_name>",} \newline
\texttt{~~"Domain description": "<domain\_desc>",} \newline
\texttt{~~"User ID": "<user\_id>",} \newline
\texttt{~~"category\_price\_level": \{"catA": "H|M|L"\}, \{"catB": "H|M|L"\},...,\{"catX": "H|M|L"\}} \newline
\texttt{~~"category\_rating\_level": \{"catA": "H|M|L"\}, \{"catB": "H|M|L"\},...,\{"catX": "H|M|L"\}} \newline
\texttt{~~"category\_popularity\_level": \{"catA": "H|M|L"\}, \{"catB": "H|M|L"\},...,\{"catX": "H|M|L"\}} \newline
\texttt{~~"category\_familiarity\_level": \{"catA": "H|M|L"\}, \{"catB": "H|M|L"\},...,\{"catX": "H|M|L"\}} \newline
\texttt{~~"overall\_category\_diversity": "H|M|L",} \newline
\texttt{~~"History": [ \{} \newline
\texttt{~~~~"item\_id": "...", "title": "...", "category": "...",} \newline
\texttt{~~~~"average\_rating": n, "rating\_number": n,} \newline
\texttt{~~~~"user\_rating": n, "cat\_item\_price\_tag": "H|M|L"} \newline
\texttt{\ \ \} ] \}} \\ \hline

\textbf{Generation Rules} & 
\begin{minipage}[t]{\linewidth}
\vspace{-0.5em}
\begin{itemize}[leftmargin=1.5em, nosep]
    \item Produce \textbf{FIVE USER PERSONA} in 3--4 sentences (at least 3).
    \item Use USER SPECIFIC INPUTS then history.
    \item Map levels to: \textbf{H}=High, \textbf{M}=Medium \textbf{L}=Low signal in <category>.
\end{itemize}

\vspace{0.5em}
\text{<NOTIONS of USER SPECIFIC INPUTS>} \newline
\scalebox{0.96}{
\begin{tabular}{@{\textbullet~~}p{\linewidth}}
    \textbf{category\_price}: Price class by category of items interacted with. \\
    \textbf{category\_rating}: Average rating level of items within each category. \\
    \textbf{category\_popularity}: Item popularity (review count) per category. \\
    \textbf{category\_familiarity}: Interaction frequency within each category. \\
    \textbf{category\_diversity}: Number of different categories explored overall.
\end{tabular}}
\vspace{0.2em}
\end{minipage} \\ \hline

\textbf{Output Format} &  
\texttt{\{ "User ID": "<user\_id>",} \newline
\texttt{~~"Profiles": \{} \newline
\texttt{~~~~"price\_sensitivity": \{ "persona": "..." \},} \newline
\texttt{~~~~"quality\_preference": \{ "persona": "..." \},} \newline
\texttt{~~~~"popularity\_bias": \{ "persona": "..." \},} \newline
\texttt{~~~~"category\_familiarity": \{ "persona": "..." \},} \newline
\texttt{~~~~"category\_diversity": \{ "persona": "..."\}} \newline\}
\\ \hline
\end{tabularx}
\end{table}

%% file: Table/table-dataset.tex
\begin{table}[t]
\centering
\caption{Dataset statistics for three Amazon domain pairs}
\vspace{-0.2cm}
\label{Table:amazon_dataset}
\resizebox{0.92\columnwidth}{!}{ 
\begin{tabular}{c|cc|cccc} 
\toprule
\textbf{Domain} & $|\mathcal{U}_o|$ & $|\mathcal{I}_d|$ & \textbf{\# Interactions} & \textbf{Train} & \textbf{Valid} & \textbf{Test} \\ \midrule
Elec & \multirow{2}{*}{7,274} & 58,461 & 180,521 & 141,868 & 17,695 & 20,958 \\
Home &  & 59,668 & 171,902 & 126,234 & 21,123 & 24,545 \\ \cline{1-7} 
Sports & \multirow{2}{*}{4,588} & 34,694 & 94,422 & 69,549 & 11,390 & 13,483 \\
Cloth &  & 45,522 & 113,708 & 82,444 & 14,553 & 16,711 \\ \cline{1-7} 
Home & \multirow{2}{*}{5,798} & 60,976 & 162,480 & 120,438 & 19,654 & 22,388 \\ 
Toys &  & 35,358 & 125,074 & 90,857 & 15,796 & 18,421 \\ \bottomrule
\end{tabular}
}
\vspace{-0.5cm}
\end{table}

%% file: Table/table-eq1.tex
\begin{table*}[t]
\vspace{-0.25cm}
\centering
\caption{Performance comparison (\%) of \ours\ on six cross-domain recommendation tasks from the Amazon dataset. 
Results on additional domain pairs and metrics are provided in Appendix.
We note that some results differ from those reported in prior work due to the time-aware data split, while all baselines are verified to be reproducible under their original settings.
}
\label{table:amazon_results}
\vspace{-0.2cm}
\resizebox{0.95\textwidth}{!}{ 
\begin{tabular}{c|c|ccc|ccccc|c} 
\toprule
\textbf{Domain} 
& \multirow{2}{*}{\textbf{\begin{tabular}[c]{@{}c@{}}Metrics\\(@5)\end{tabular}}}
& \multicolumn{3}{c|}{\textbf{Single-Domain Methods}} 
& \multicolumn{5}{c|}{\textbf{Cross-Domain Methods}} 
& \multirow{2}{*}{\textbf{\ours}} \\ 
\textbf{(Source $\to$ Target)} 
& 
& BPRMF & NGCF & LightGCN
& UniCDR & CDIMF & CUT-LightGCN & PPA & DGCDR
&  \\ \midrule

\multirow{2}{*}{\begin{tabular}[c]{@{}c@{}}Elec $\to$ Home\end{tabular}} 
& HR    & 1.69±0.10 & 1.66±0.13 & 1.77±0.10 & 1.69±0.03 & 1.94±0.06 & 1.96±0.08 & \underline{3.06±0.04} & 2.63±0.12 & \textbf{3.23±0.19} \\
& NDCG  & 0.49±0.03 & 0.49±0.04 & 0.51±0.02 & 0.54±0.01 & 0.57±0.02 & 0.58±0.03 & \underline{0.89±0.01} & 0.76±0.03 & \textbf{0.91±0.03} \\
\hline

\multirow{2}{*}{\begin{tabular}[c]{@{}c@{}}Home $\to$ Elec\end{tabular}} 
& HR    & 1.66±0.15 & 1.48±0.08 & 2.21±0.10 & 1.55±0.08 & 1.64±0.06 & 1.57±0.05 & \underline{1.79±0.01} & 1.57±0.08 & \textbf{2.44±0.38} \\
& NDCG  & 0.47±0.02 & 0.42±0.02 & 0.63±0.03 & 0.47±0.02 & 0.46±0.01 & 0.46±0.01 & \underline{0.54±0.01} & 0.47±0.02 & \textbf{0.77±0.15} \\
\midrule

\multirow{2}{*}{\begin{tabular}[c]{@{}c@{}}Sports $\to$ Cloth\end{tabular}}
& HR    & 1.56±0.36 & 1.51±0.23 & 1.66±0.14 & 1.71±0.23 & 2.21±0.05 & 2.96±0.18 & \underline{3.36±0.05} & 2.25±0.17 & \textbf{4.09±0.58} \\
& NDCG  & 0.42±0.07 & 0.42±0.07 & 0.46±0.02 & 0.47±0.06 & 0.68±0.01 & 0.92±0.08 & \underline{0.98±0.01} & 0.63±0.04 & \textbf{1.11±0.31} \\
\hline

\multirow{2}{*}{\begin{tabular}[c]{@{}c@{}}Cloth $\to$ Sports\end{tabular}} 
& HR    & 1.59±0.20 & 1.82±0.12 & 2.20±0.14 & 1.37±0.10 & 1.37±0.05 & 1.67±0.06 & \textbf{2.25±0.04} & 1.85±0.05 & \underline{2.08±0.19} \\
& NDCG  & 0.50±0.06 & 0.56±0.03 & 0.72±0.04 & 0.42±0.05 & 0.39±0.02 & 0.51±0.01 & \textbf{0.71±0.01} & 0.58±0.01 & \underline{0.68±0.10} \\
\midrule
\multirow{2}{*}{\begin{tabular}[c]{@{}c@{}}Home $\to$ Toys\end{tabular}} 
& HR    & 0.95±0.11 & 0.84±0.09 & 0.96±0.11 & 0.91±0.07 & 1.31±0.01 & 1.23±0.08 & \underline{1.35±0.05} & 1.19±0.04 & \textbf{1.56±0.24} \\
& NDCG  & 0.27±0.01 & 0.23±0.03 & 0.25±0.02 & 0.27±0.02 & 0.41±0.00 & 0.37±0.02 & \underline{0.36±0.02} & 0.30±0.01 & \textbf{0.47±0.03} \\
\hline

\multirow{2}{*}{\begin{tabular}[c]{@{}c@{}}Toys $\to$ Home\end{tabular}} 
& HR    & 1.44±0.28 & 0.88±0.14 & 1.75±0.11 & 1.85±0.14 & 1.55±0.02 & 2.13±0.08 & \underline{2.98±0.04} & 2.73±0.06 & \textbf{3.11±0.15} \\
& NDCG  & 0.40±0.07 & 0.23±0.04 & 0.47±0.04 & 0.51±0.04 & 0.43±0.00 & 0.58±0.03 & \underline{0.80±0.01} & 0.73±0.02 & \textbf{0.88±0.03} \\
\bottomrule
\end{tabular}
}
\vspace{-0.1cm}
\end{table*}

%% file: Sections/Eval.tex
To evaluate the effectiveness of \ours, we designed our experiments to answer the following  evaluation questions (EQs): 
\begin{itemize} [leftmargin=*]
    \item \textbf{(EQ1)}: Does {\ours} consistently outperform state-of-the-art CDR methods on real-world benchmarks?
    \item \textbf{(EQ2)}: Does domain-specific persona modeling effectively capture fine-grained user preferences for CDR?
    \item \textbf{(EQ3)}: Does doppelganger-based knowledge transfer provide more reliable cross-domain knowledge transfer?
    \item \textbf{(EQ4)}: How sensitive is the accuracy of \ours\ to variations in the hyper-parameters $\lambda$ and $\tau$?
    \item \textbf{(EQ5)}: Is \ours\ computationally efficient in practice?
\end{itemize}

In addition, we conduct supplementary analyses using open-source LLM variants and a no-LLM variant to further examine the robustness and necessity of the LLM-assisted persona modeling component in \ours.
Detailed results are provided in \textbf{Appendix~\ref{app_analysis_n_result}} (see Table~\ref{tab:app-openLLM} and Table~\ref{tab:app-noLLM}).

\subsection{Experimental Setup}
\noindent\textbf{Datasets.}
We evaluate {\ours} on five domains from the widely used Amazon review dataset \cite{amazondata}, including Electronics, Home, Sports, Cloth, and Toys.
All datasets are publicly available.\footnote{https://amazon-reviews-2023.github.io/.}
We conduct experiments on all possible cross-domain pairs, resulting in ten domain transfer settings.
In this section, we focus on three commonly used domain pairs with the largest numbers of overlapping users.
Table~\ref{Table:amazon_dataset} reports statistics for the three domain pairs.

The Amazon dataset covers user interactions from 1996 to 2023.
Unlike random-sampling-based protocols adopted in prior work \cite{CDRIB, SSCDR, UniCDR, zhang2024ppa}, we employ a time-aware data split for temporal consistency~\cite{lee2025caper}.
Specifically, interactions before 2019 are used as the training set, while interactions after 2019 are split into validation and test sets, considering only users observed in both periods.

\vspace{1mm}
\noindent\textbf{Evaluation Tasks.}
Following prior CDR studies \cite{cut, ju2024drip, cdcdr}, we adopt a \textit{full ranking} evaluation protocol, where all unobserved items are treated as candidates for each user. 
We evaluate performance using standard top-$K$ metrics, including hit ratio (HR) and normalized discounted cumulative gain (NDCG). 
All results are averaged over \textit{five} independent runs with different random seeds.


\input{Table/table-eq2-1.tex}

\vspace{1mm}
\noindent\textbf{Competitors.}
We compare \ours\ with eight baselines under two settings: single-domain recommendation methods, including BPRMF~\cite{BPR}, NGCF~\cite{wang2019ngcf}, and LightGCN~\cite{he2020lightgcn}, and cross-domain recommendation methods, including UniCDR~\cite{UniCDR}, CDIMF~\cite{samra2024cdimf}, CUT~\cite{cut}, PPA~\cite{zhang2024ppa}, and DGCDR~\cite{wang2025dgcdr}.
Detailed descriptions of the CDR baselines are provided in Section~\ref{sec:rworks}.
For evaluation, we use implementations from the widely adopted open-source library QRec\footnote{https://github.com/Coder-Yu/QRec.}, and resort to author-released code when unavailable. 

\vspace{1mm}
\noindent\textbf{Implementation details.}
All experiments are conducted using PyTorch on NVIDIA RTX 6000 Ada GPUs.
Following \cite{yang2024not, cdcdr, BiTGCF}, we use the Adam optimizer \cite{kingma2014adam} and train all models for up to 100 epochs with early stopping.
We fix the batch size to 1024 and set the embedding dimension for users and items to 128. 
The dropout rate is tuned over \{0.1, 0.2, 0.3, 0.4, 0.5, 0.6, 0.7\}, the number of negative samples is selected from \{1, 3, 5\}, and the regularization coefficient is searched over \{1e-4, 5e-4, 1e-3, 5e-3\}. 
All hyperparameters are tuned on the validation set. 

For persona construction in \ours, we use \texttt{gpt-4o} with each user's most recent 30 interactions, resulting in an average per-user cost of  approximately \$0.022. 
The temperatures for persona generation and domain description generation are set to $0.7$ and $1.0$ separately.
For semantic encoding, we employ \texttt{text-embedding-3-large}, using $3,072$-dimensional embeddings for user personas and 512-dimensional embeddings for item semantic representations.

\input{Table/table-eq2-2.tex}

\subsection{Results and Analysis}\label{sec:result}
Due to space limitations, we report results for three representative domain pairs using top-5 metrics, and provide complete results on other domain pairs and metrics in \textbf{Appendix~\ref{app_analysis_n_result}}.

\vspace{1mm}
\noindent{\bf(EQ1) Overall Performance Comparison.}
Table~\ref{table:amazon_results} reports the overall performance of \ours\ compared with eight baselines. 
\textbf{Bold} and \uline{underlined} values indicate the best and second-best results, respectively.
As shown in Table~\ref{table:amazon_results}, \ours\ achieves the best performance in 5 out of 6 bidirectional CDR tasks, demonstrating consistent superiority across most transfer settings. 
Specifically, on the (Home$\rightarrow$Elec) task, \ours\ outperforms the best competitor (\ie, PPA) by up to 36.3\%, and 42.6\% in terms of HR@5 and NDCG@5, respectively.
The only exception occurs in the (Cloth$\rightarrow$Sports) task, where \ours\ ranks second-best; however, it significantly outperforms the same baseline (\ie, PPA) in the reverse direction. 
Overall, these results demonstrate that explicitly modeling intra-domain heterogeneity through multi-persona representations is crucial for achieving robust and consistent CDR performance.




\vspace{1mm}
\noindent{\bf (EQ2) Effectiveness of Persona Modeling.} 
To examine the effectiveness of persona modeling in \ours, we evaluate three aspects: persona decomposition with processed metadata (\textbf{EQ2-1}), context-aware aggregation of decomposed personas (\textbf{EQ2-2}), and robustness under cold-start target-domain scenarios (\textbf{EQ2-3}).

For (EQ2-1), Table \ref{tab:eq2-1} compares three settings\footnote{Prompts and sample results for these variants are provided in the \href{https://github.com/archivehee/Multi-TAP/tree/main/Multi-TAP/system_prompt}{\textbf{\uline{online appendix}}}.}: a \textit{single}-persona model using \textit{raw} item-level metadata (Single (Raw)), a \textit{single}-persona model using \textit{processed} metadata (Single (Proc.)), and our proposed \textit{multi}-persona model using the same \textit{processed} metadata (Multi (Proc.)).
Here, raw metadata provides criterion-specific values directly for each preference dimension, whereas processed metadata encodes their relative ordinal levels, denoted as $l^{\mathrm{PS}}$, $l^{\mathrm{QP}}$, $l^{\mathrm{PB}}$, $l^{\mathrm{CF}}$, and $l^{\mathrm{CD}}$.
We make two key observations from Table~\ref{tab:eq2-1}.
First, replacing raw item-level metadata with processed metadata consistently improves performance, indicating that relative ordinal preference signals provide more informative supervision than absolute criterion values.
Second, even under the same processed metadata, multi-persona decomposition yields additional gains over single-persona modeling, demonstrating the importance of explicitly modeling intra-domain heterogeneity.

For (EQ2-2), Table~\ref{tab:eq2-2} compares different aggregation strategies for integrating decomposed personas.
Specifically, we consider two simple baselines, \textit{Mean} and \textit{Concat}, which aggregate persona representations via uniform averaging and feature concatenation, respectively.
We compare them against \textit{Self-Attn}, our proposed self-attentive aggregation module, which adaptively weights personas conditioned on the user and the target domain.
As shown in Table~\ref{tab:eq2-2}, \textit{Self-Attn} consistently outperforms \textit{Mean} and \textit{Concat}, indicating that adaptive self-attentive weighting is crucial for effectively integrating decomposed user personas.

\input{Table/table-eq2-3.tex}

For (EQ2-3), Table~\ref{tab:eq2-3} evaluates the robustness of \ours\ under cold-start target-domain scenarios. 
Following prior CDR protocols~\cite{BiTGCF, yang2024not}, we randomly sample 10\% of users and restrict their target-domain interactions to fewer than five in the training data. 
Even under this challenging condition, \ours\ consistently outperforms the strong baseline PPA \cite{zhang2024ppa}. 
In particular, \ours\ improves HR@5 and NDCG@5 by up to 14.52\% and 25.81\%, respectively. 
These results suggest that decomposed persona representations enable more robust cross-domain preference transfer under sparse interaction conditions.

\vspace{1mm}
\noindent{\bf (EQ3) Effectiveness of Doppelganger Transfer.} 
We evaluate the effectiveness of our target-adaptive doppelganger transfer by comparing it with \textit{direct} InfoNCE-based alignment strategies, including 
(1) \textit{Direct-ID}, which aligns ID-based user embeddings $\mathbf{v}_{u_s}$ and $\mathbf{v}_{u_t}$ from LightGCN, 
(2) \textit{Direct-Persona}, which aligns persona embeddings $\mathbf{h}_{u_s}$ and $\mathbf{h}_{u_t}$, and 
(3) \textit{Direct-Both}, which aligns their concatenation.
All variants share the same backbone as \ours\ and differ only in the transfer mechanism.
As shown in Figure~\ref{fig:transfer_strategy}, \ours\ consistently outperforms all direct alignment baselines, while direct alignment exhibits unstable performance across transfer directions.
This indicates that enforcing direct representation alignment can be sensitive to domain characteristics and may suppress domain-specific preference signals.

\input{Plot/plot-eq3.tex}

\vspace{1mm}
\noindent{\bf (EQ4) Hyperparameter Sensitivity.} 
We examine the sensitivity of \ours\ to two hyperparameters: the weight $\lambda$ of the doppelganger-based transfer loss and the temperature $\tau$.
Figure~\ref{fig:lambda_tau_sensitivity} illustrates the performance trends.
For $\lambda$, non-trivial values consistently improves performance, with larger values (\eg, $>1$) achieving clear gains over small ones, highlighting the importance of semantic knowledge transfer beyond behavioral learning.
For $\tau$, performance is stable across domain pairs, with optimal results typically observed in the range of $[0.4, 0.6]$, indicating that a moderate temperature is sufficient for stable doppelganger-based alignment.



\vspace{1mm}
\noindent{\bf (EQ5) Computational Efficiency.} 
We examine the computational efficiency of \ours\ by measuring its training cost.
Under the configurations used in our experiments, \ours\ can be trained within 9–30 minutes across different domain pairs.
This efficiency stems from building on the backbone embeddings learned by LightGCN and introducing only lightweight persona aggregation and contrastive components.
Persona generation incurs an average cost of about 3 seconds per user and is performed offline, ensuring that the overall training pipeline remains scalable.


%% file: Table/table-eq2-1.tex
\begin{table}[t]
\centering
\caption{Effect of persona decomposition with processed metadata, where Multi (Proc.) denotes our strategy}
\label{tab:eq2-1}
\vspace{-0.25cm}
\resizebox{1\columnwidth}{!}{ 
\begin{tabular}{cc|ccc} 
\toprule
\textbf{Domain} 
& \textbf{Metrics (@5)} 
& \textbf{Single (Raw)}
& \textbf{Single (Proc.)}
& \textbf{Multi (Proc.)} \\
\midrule

\multirow{2}{*}{Sports $\rightarrow$ Cloth}
& HR    & 3.70 & \underline{3.77} & \textbf{4.09} \\
& NDCG  & 1.08 & \underline{1.09} & \textbf{1.11} \\
\midrule

\multirow{2}{*}{Cloth $\rightarrow$ Sports}
& HR    & 1.83 & \underline{1.99} & \textbf{2.08} \\
& NDCG  & 0.61 & \underline{0.64} & \textbf{0.68} \\
\bottomrule
\end{tabular}
}
\vspace{-0.2cm}
\end{table}

%% file: Table/table-eq2-2.tex
\begin{table}[t]
\centering
\caption{Effect of self-attentive aggregation for decomposed personas, where Self-Attn denotes our strategy}
\label{tab:eq2-2}
\vspace{-0.25cm}
\resizebox{0.9\columnwidth}{!}{ 
\begin{tabular}{cc|ccc} 
\toprule
\textbf{Domain} 
& \textbf{Metrics (@5)} 
& \textbf{Mean}
& \textbf{Concat}
& \textbf{Self-Attn} \\ 
\midrule

\multirow{2}{*}{Sports $\rightarrow$ Cloth} 
& HR   & 3.66 & \underline{3.99} & \textbf{4.09} \\
& NDCG & \underline{1.10} & 1.04 & \textbf{1.11} \\ 
\midrule

\multirow{2}{*}{Cloth $\rightarrow$ Sports} 
& HR   & 1.93 & \underline{2.01} & \textbf{2.08} \\
& NDCG & 0.60 & \underline{0.65} & \textbf{0.68} \\ 
\bottomrule
\end{tabular}
}
\vspace{-0.4cm}
\end{table}

%% file: Table/table-eq2-3.tex
\begin{table}[t]
\centering
\caption{Effect of \ours\ under cold-start scenarios}
\label{tab:eq2-3}
\vspace{-0.25cm}
\resizebox{0.8\columnwidth}{!}{ 
\begin{tabular}{cc|cc} 
\toprule
\textbf{Domain} 
& \textbf{Metrics (@5)} 
& \textbf{PPA}
& \textbf{Multi-TAP} \\ 
\midrule

\multirow{2}{*}{Elec $\rightarrow$ Home} 
& HR   & \textbf{3.25 $\pm$ 0.29} & \underline{3.19 $\pm$ 0.31} \\
& NDCG & \underline{0.80 $\pm$ 0.03} & \textbf{0.82 $\pm$ 0.07} \\ 
\midrule

\multirow{2}{*}{Home $\rightarrow$ Elec} 
& HR   & \underline{2.31 $\pm$ 0.02} & \textbf{2.42 $\pm$ 0.45} \\
& NDCG & \underline{0.69 $\pm$ 0.07} & \textbf{0.75 $\pm$ 0.13} \\ 
\midrule

\multirow{2}{*}{Home $\rightarrow$ Toys} 
& HR   & \underline{1.24 $\pm$ 0.32} & \textbf{1.42 $\pm$ 0.29} \\
& NDCG & \underline{0.31 $\pm$ 0.07} & \textbf{0.39 $\pm$ 0.13} \\ 
\midrule

\multirow{2}{*}{Toys $\rightarrow$ Home} 
& HR   & \textbf{2.26 $\pm$ 0.11} & \underline{2.20 $\pm$ 0.27} \\
& NDCG & \underline{0.53 $\pm$ 0.03} & \textbf{0.55 $\pm$ 0.07} \\ 
\bottomrule
\end{tabular}
}
\vspace{-0.5cm}
\end{table}

%% file: Plot/plot-eq3.tex
\begin{figure}[t]
\footnotesize
\centering
\begin{tikzpicture}
\begin{groupplot}[
    group style={
        group size=3 by 1,
        horizontal sep=0.2cm, 
        xlabels at=edge bottom,
        y descriptions at=edge left,
    },
    width=3.7cm,  
    height=3.5cm, 
    ybar=0pt,     
    enlarge x limits=0.7,
    symbolic x coords={pairs},
    xtick=\empty,
    ymin=0,
    ymax=0.045,
    ymajorgrids=true,
    title style={font=\bfseries\tiny, yshift=-3pt, align=center},
    ylabel={HR@5},
    ylabel style={font=\footnotesize, xshift=6pt},
    yticklabel style={
        font=\footnotesize,
        /pgf/number format/fixed,
        /pgf/number format/precision=3
    },
    scaled y ticks=false,
]

\nextgroupplot[title={Elec \hspace{0.5cm} Home \\ $\to$ Home \hspace{0.3cm} $\rightarrow$ Elec}, bar width=5pt]
\addplot[fill=blue!30, bar shift=-20.0pt] coordinates {(pairs, 0.0283)};
\addplot[fill=red!30,  bar shift=-15.0pt] coordinates {(pairs, 0.0275)};
\addplot[fill=gray!45, bar shift=-10.0pt] coordinates {(pairs, 0.0275)};
\addplot[fill=orange,  bar shift=-5.0pt]  coordinates {(pairs, 0.0323)};
\addplot[fill=blue!30, bar shift=5.0pt]  coordinates {(pairs, 0.0211)};
\addplot[fill=red!30,  bar shift=10.0pt]  coordinates {(pairs, 0.0206)};
\addplot[fill=gray!45, bar shift=15.0pt] coordinates {(pairs, 0.0188)};
\addplot[fill=orange,  bar shift=20.0pt] coordinates {(pairs, 0.0244)};

\nextgroupplot[title={Sports \hspace{0.5cm} Cloth \\ $\to$ Cloth \hspace{0.3cm} $\rightarrow$ Sports},bar width=5pt, ylabel={}]
\addplot[fill=blue!30, bar shift=-20.0pt] coordinates {(pairs, 0.0259)};
\addplot[fill=red!30,  bar shift=-15pt] coordinates {(pairs, 0.0346)};
\addplot[fill=gray!45, bar shift=-10pt] coordinates {(pairs, 0.0363)};
\addplot[fill=orange,  bar shift=-5pt]  coordinates {(pairs, 0.0409)};
\addplot[fill=blue!30, bar shift=5pt]  coordinates {(pairs, 0.0188)};
\addplot[fill=red!30,  bar shift=10pt]  coordinates {(pairs, 0.0192)};
\addplot[fill=gray!45, bar shift=15pt] coordinates {(pairs, 0.0193)};
\addplot[fill=orange,  bar shift=20pt] coordinates {(pairs, 0.0208)};

\nextgroupplot[title={Home \hspace{0.5cm} Toys \\ $\to$ Toys \hspace{0.3cm} $\rightarrow$ Home},bar width=5pt, ylabel={}]
\addplot[fill=blue!30, bar shift=-20.0pt] coordinates {(pairs, 0.0113)};
\addplot[fill=red!30,  bar shift=-15.pt] coordinates {(pairs, 0.0136)};
\addplot[fill=gray!45, bar shift=-10pt] coordinates {(pairs, 0.0121)};
\addplot[fill=orange,  bar shift=-5pt]  coordinates {(pairs, 0.0156)};
\addplot[fill=blue!30, bar shift=5pt]  coordinates {(pairs, 0.0275)};
\addplot[fill=red!30,  bar shift=10pt]  coordinates {(pairs, 0.0272)};
\addplot[fill=gray!45, bar shift=15.0pt] coordinates {(pairs, 0.0301)};
\addplot[fill=orange,  bar shift=20pt] coordinates {(pairs, 0.0311)};

\end{groupplot}
\end{tikzpicture}

\vspace{-0.2cm}
\begin{tikzpicture}
\centering
\matrix[column sep=0.8ex, row sep=0pt] {
\node[fill=blue!30, minimum width=3pt, minimum height=3pt]{}; & \node{\scalebox{1}{Direct-ID}}; &
\node[fill=red!30,  minimum width=3pt, minimum height=3pt]{}; & \node{\scalebox{1}{Direct-Persona}}; &
\node[fill=gray!45, minimum width=3pt, minimum height=3pt]{}; & \node{\scalebox{1}{Direct-Both}}; &
\node[fill=orange,  minimum width=3pt, minimum height=3pt]{}; & \node{\scalebox{1}{\ours}}; \\
};
\end{tikzpicture}
\vspace{-0.25cm}
\caption{Effect of doppelganger-based indirect transfer compared to direct alignment strategies in terms of HR@5.}
\label{fig:transfer_strategy}
\vspace{-0.4cm}
\end{figure}
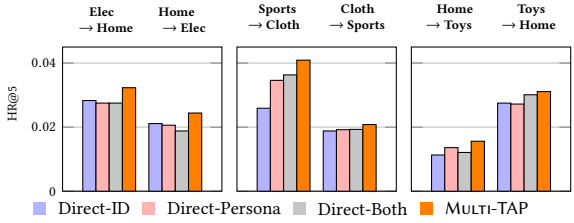

%% file: Sections/RWorks.tex

Early CDR work \cite{EMCDR,SSCDR,DTCDR,GA-DTCDR} focuses on \textit{representation alignment} using overlapping users, where mapping functions project source-domain user embeddings into the target space, as in EMCDR \cite{EMCDR} and CDIMF \cite{samra2024cdimf}.
Subsequent studies \cite{CAT-ART, UniCDR, DisenCDR} distinguish \textit{domain-shared} and \textit{domain-specific} preferences via decomposition or contrastive learning, with representative methods including UniCDR \cite{UniCDR} and EDDA \cite{EDDA}.
More recent studies address \textit{negative transfer mitigation}  by filtering irrelevant source-domain signals, such as CUT \cite{cut} and TrineCDR \cite{song2024trinecdr}, or by revisiting direct alignment through diffusion-based reconstruction, as in DGCDR \cite{wang2025dgcdr}.
However, these methods still operate primarily at the domain level and do not explicitly model intra-domain preference heterogeneity.

A few CDR studies explore modeling user preferences with multiple latent components \cite{sun2023remit,li2025disco, zhang2024ppa}.
PPA \cite{zhang2024ppa} introduces \textit{latent preference prototypes} learned solely from interaction data to capture heterogeneous user interests within ID-based representations.
However, these prototypes are implicitly inferred, their number is treated as a hyperparameter, and they are not explicitly adapted to the target domain during transfer.
Meanwhile, LLM-based approaches, such as LASSO \cite{chen2025lasso}, construct semantic user profiles and align users across domains at the natural language level.
However, these methods often rely on implicit reasoning to compress rich metadata into a \textit{single, unstructured user profile}, which limits preference controllability and raises concerns about transfer stability.

%% file: Sections/Conclusion.tex
\input{Plot/plot-eq4.tex}

In this work, we identified that most existing cross-domain recommendation methods fail to explicitly capture \textit{intra-domain preference heterogeneity}, which is crucial for stable and effective knowledge transfer.
Motivated by this observation, we proposed \ours, a multi-persona CDR framework that models heterogeneous user preferences in a structured manner and performs target-adaptive knowledge transfer via doppelganger alignment.
Extensive experiments show that \ours\ consistently outperforms state-of-the-art methods while remaining computationally efficient.

%% file: Plot/plot-eq4.tex
\begin{figure}[t]
\small
\centering
\vspace{0.05cm}

\begin{tikzpicture}
\begin{groupplot}[
    group style={
        group size=2 by 1,
        horizontal sep=0.55cm,
        ylabels at=edge left,
        xlabels at=edge bottom
    },
    width=5cm,
    height=3.5cm,
    ymin=0,
    ymajorgrids=true,
    grid style={gray!30},
    xlabel style={font=\footnotesize},
    ylabel style={font=\footnotesize},
    tick label style={font=\footnotesize},
    legend style={draw=none},
    scaled y ticks=false,
    yticklabel style={
    /pgf/number format/fixed,
    /pgf/number format/precision=3
},
    y tick scale label code/.code={},
    xticklabel style={/pgf/number format/fixed,/pgf/number format/precision=1},
]

\nextgroupplot[
    xlabel={$\lambda$},
    ylabel={HR@5},
    xmin=0, xmax=2.0,
    xtick={0,0.2, 0.4, 0.6, 0.8, 1.0, 1.2, 1.4, 1.6, 1.8, 2.0},
]

\addplot[
    thick, color=blue,
    mark=o, mark size=1.5pt,
] coordinates {
    (0.0,0.0176) (0.2,0.0227) (0.4,0.0232) (0.6,0.0228) (0.8,0.0249)
    (1.0,0.0258) (1.2,0.0270) (1.4,0.0320) (1.6,0.0291) (1.8,0.0287) (2.0,0.0291)
};

\addplot[
    thick, color=red,
    mark=triangle*, mark size=1.5pt,
] coordinates {
    (0.0,0.0153) (0.2,0.0206) (0.4,0.0212) (0.6,0.0211) (0.8,0.0223)
    (1.0,0.0221) (1.2,0.0234) (1.4,0.0244) (1.6,0.0234) (1.8,0.0229) (2.0,0.0231)
};

\addplot[
    thick, color=orange!85!black,
    mark=square*, mark size=1.5pt,
] coordinates {
    (0.0,0.0235) (0.2,0.0292) (0.4,0.0318) (0.6,0.0336) (0.8,0.0355)
    (1.0,0.0372) (1.2,0.0386) (1.4,0.0397) (1.6,0.0401) (1.8,0.0405) (2.0,0.0398)
};

\addplot[
    thick, color=green!55!black,
    mark=star,
    mark options={fill=green!55!black},
    mark size=1.8pt,
] coordinates {
    (0.0,0.0112) (0.2,0.0142) (0.4,0.0158) (0.6,0.0172) (0.8,0.0184)
    (1.0,0.0193) (1.2,0.0207) (1.4,0.0210) (1.6,0.0204) (1.8,0.0198) (2.0,0.0195)
};

\addplot[
    thick, color=purple!75!black,
    mark=diamond*,
    mark options={fill=purple!75!black},
    mark size=1.8pt,
] coordinates {
    (0.0,0.0058) (0.2,0.0094) (0.4,0.0103) (0.6,0.0112) (0.8,0.0121)
    (1.0,0.0127) (1.2,0.0135) (1.4,0.0154) (1.6,0.0144) (1.8,0.0147) (2.0,0.0129)
};

\addplot[
    thick, color=teal!70!black,
    mark=x,
    mark options={fill=teal!70!black},
    mark size=1.7pt,
] coordinates {
    (0.0,0.0195) (0.2,0.0230) (0.4,0.0255) (0.6,0.0272) (0.8,0.0286)
    (1.0,0.0297) (1.2,0.0310) (1.4,0.0305) (1.6,0.0296) (1.8,0.0289) (2.0,0.0283)
};

\nextgroupplot[
    xlabel={$\tau$},
    xmin=0.1, xmax=1.0,
    xtick={0,0.1, 0.2, 0.3, 0.4, 0.5, 0.6, 0.7, 0.8, 0.9, 1},
]

\addplot[thick, color=blue, mark=o, mark size=1.5pt] coordinates {
    (0.1,0.0207) (0.2,0.0203) (0.3,0.0211) (0.4,0.0297) (0.5,0.0297)
    (0.6,0.0300) (0.7,0.0239) (0.8,0.0212) (0.9,0.0210) (1.0,0.0205)
};

\addplot[thick, color=red, mark=triangle*, mark size=1.5pt] coordinates {
    (0.1,0.0211) (0.2,0.0218) (0.3,0.0230) (0.4,0.0239) (0.5,0.0244)
    (0.6,0.0243) (0.7,0.0241) (0.8,0.0234) (0.9,0.0227) (1.0,0.0221)
};

\addplot[thick, color=orange!85!black, mark=square*, mark size=1.5pt] coordinates {
    (0.1,0.0378) (0.2,0.0386) (0.3,0.0397) (0.4,0.0406) (0.5,0.0409)
    (0.6,0.0408) (0.7,0.0406) (0.8,0.0398) (0.9,0.0390) (1.0,0.0383)
};

\addplot[
    thick, color=green!55!black,
    mark=star, mark options={fill=green!55!black},
    mark size=1.8pt,
] coordinates {
    (0.1,0.0188) (0.2,0.0194) (0.3,0.0201) (0.4,0.0206) (0.5,0.0208)
    (0.6,0.0207) (0.7,0.0206) (0.8,0.0201) (0.9,0.0196) (1.0,0.0192)
};

\addplot[
    thick, color=purple!75!black,
    mark=diamond*, mark options={fill=purple!75!black},
    mark size=1.8pt,
] coordinates {
    (0.1,0.0128) (0.2,0.0138) (0.3,0.0156) (0.4,0.0154) (0.5,0.0153)
    (0.6,0.0152) (0.7,0.0150) (0.8,0.0144) (0.9,0.0139) (1.0,0.0134)
};

\addplot[
    thick, color=teal!70!black,
    mark=x, mark options={fill=teal!70!black},
    mark size=1.8pt,
] coordinates {
    (0.1,0.0276) (0.2,0.0293) (0.3,0.0311) (0.4,0.0308) (0.5,0.0306)
    (0.6,0.0304) (0.7,0.0301) (0.8,0.0292) (0.9,0.0284) (1.0,0.0279)
};

\end{groupplot}
\end{tikzpicture}

\vspace{-0.05cm}
\begin{tikzpicture}
\matrix[column sep=1.4ex, row sep=0.2ex] {

\node[fill=blue,   minimum width=4pt, minimum height=4pt, inner sep=0pt]{}; & \node{\footnotesize Elec$\to$Home}; &
\node[fill=orange!85!black, minimum width=4pt, minimum height=4pt, inner sep=0pt]{}; & \node{\footnotesize Sports$\to$Cloth}; & 
\node[fill=purple!75!black, minimum width=4pt, minimum height=4pt, inner sep=0pt]{}; & \node{\footnotesize Home$\to$Toys}; \\
\node[fill=red,    minimum width=4pt, minimum height=4pt, inner sep=0pt]{}; & \node{\footnotesize Home$\to$Elec}; &
\node[fill=green!55!black, minimum width=4pt, minimum height=4pt, inner sep=0pt]{}; & \node{\footnotesize Cloth$\to$Sports}; &
\node[fill=teal!70!black, minimum width=4pt, minimum height=4pt, inner sep=0pt]{}; & \node{\footnotesize Toys$\to$Home}; \\
};

\end{tikzpicture}
\vspace{-0.25cm}
\caption{Sensitivity of \ours\ to the hyperparameters $\boldsymbol{\lambda}$ and $\boldsymbol{\tau}$ in terms of HR@5.}
\label{fig:lambda_tau_sensitivity}
\vspace{-0.25cm}
\end{figure}
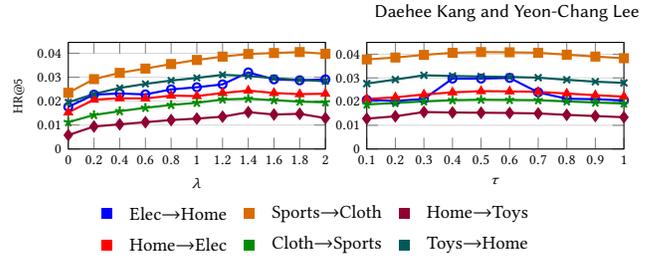

%% file: Sections/Appendix.tex
\section{Additional Details of Multi-TAP}~\label{app} 


\vspace{-5mm}
\subsection{User Persona Database}~\label{app_persona_db} 

\vspace{-2mm}
\noindent \textbf{Table~\ref{tab:user_persona_db}} presents a sample from the pre-built user persona database used as structured input for multi-persona construction.

\vspace{-2mm}
\input{Table/table-persona_db.tex}

\vspace{-2mm}
\subsection{LLM-based Domain Description Generation}~\label{app_sysprompt} 

\vspace{-2mm}
\noindent We derive a compact domain description using an LLM to summarize domain-specific item characteristics. 
This description provides contextual guidance that constrains user persona generation $f_u^{k}$ within appropriate domain-specific semantic space. 
To construct the description for each domain $d$, we sample representative items based on interaction frequency and rating scores. 
Specifically, we select 50 frequently interacted items and 50 highly rated items, resulting in 100 representative items per domain.
Using their associated metadata, we construct domain configuration texts $\mathcal{T}_{i,d}$:
\begin{equation}
\mathcal{T}_{i,d} = (\text{Domain} \,\Vert\, \text{Item title} \,\Vert\, \text{Item category} \,\Vert\, \text{Item description}),
\end{equation}
where $\Vert$ denotes text concatenation. 
The domain description is then generated as follows:
\begin{equation}
\mathcal{T}^{\text{desc}}_d = \text{LLM}\bigl(\mathbb{P}^{\text{desc}} \,\Vert\, \mathcal{T}_{1,d} \,\Vert\, \cdots \,\Vert\, \mathcal{T}_{100,d} \bigr),
\end{equation}
where $\mathbb{P}^{\text{desc}}$ denotes the system prompt for domain description generation and $\text{LLM}(\cdot)$ represents the text generation process of the language model. 
The system prompt and an example output are provided in \textbf{Table~\ref{tab:domain_description_prompt}}.



\vspace{-2mm}
\section{Additional Analysis and Results}~\label{app_analysis_n_result} 

\vspace{-2mm}
\noindent Following the main experimental results, we present additional analyses to further validate the effectiveness of \ours.

\vspace{1mm}
\noindent\textbf{Robust and necessity of LLM-assisted persona modeling.}
\textbf{Table~\ref{tab:app-openLLM}} and \textbf{Table~\ref{tab:app-noLLM}} present additional analyses using open-source LLM variants and a no-LLM variant of \ours.
The results show that \ours\ maintains competitive performance across different LLM backbones, while consistently outperforming the no-LLM variant.
These findings suggest that \ours\ is robust across different LLM backbones and that LLM-generated personas provide benefits beyond simple structured features.

\vspace{1mm}
\noindent\textbf{Complete performance comparison.}
\textbf{Table~\ref{table:app_amazon_results}} reports complete performance results for the remaining cross-domain pairs in the Amazon dataset. 
Across diverse transfer scenarios, \ours\ consistently outperforms state-of-the-art baselines, demonstrating the robustness of our multi-persona decomposition and doppelganger-based transfer strategy.

\begin{figure*}[t]
\centering
\vspace{1cm}
\includegraphics[width=0.95\textwidth]{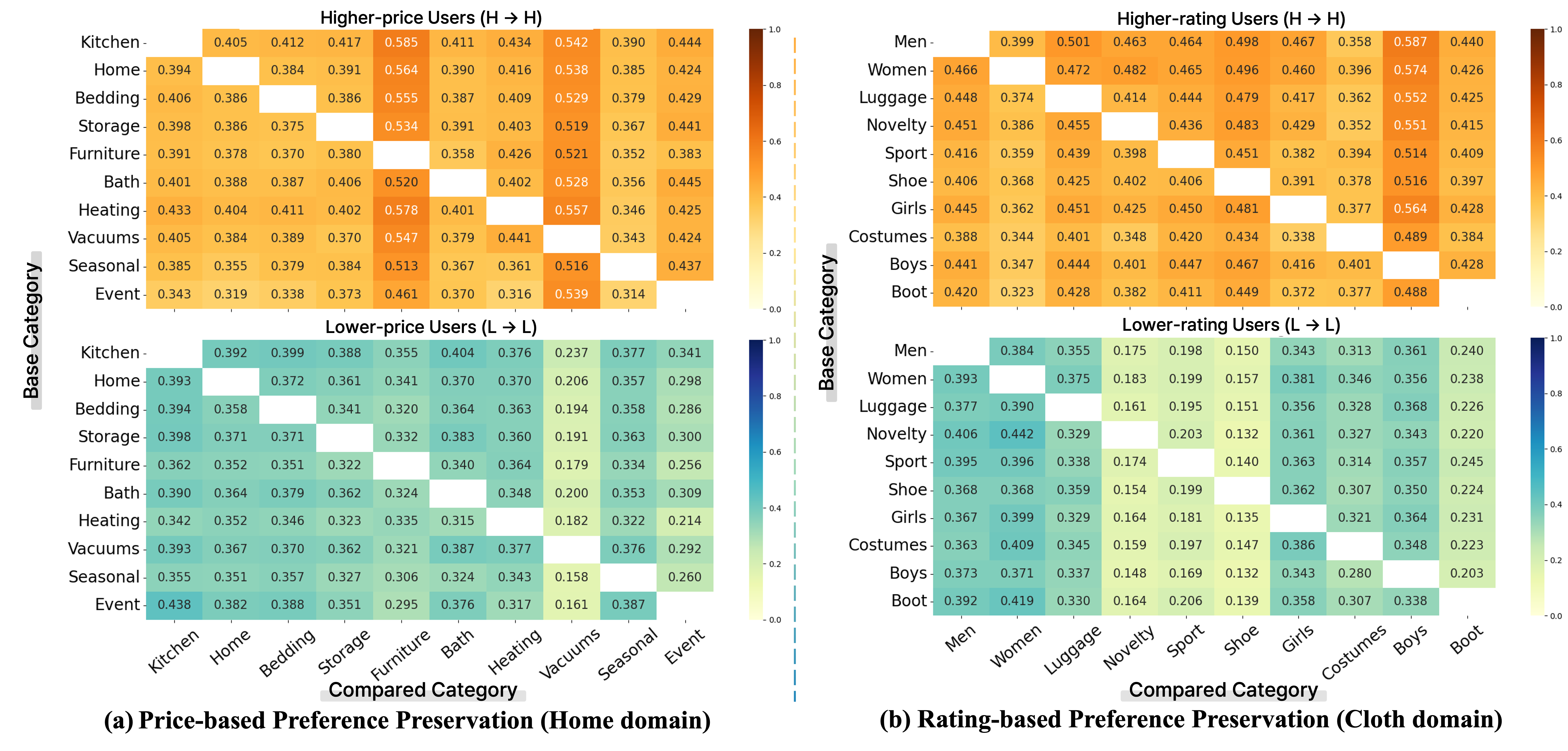}



\vspace{-0.3cm} 
\caption{Additional analysis of intra-domain category-level preference heterogeneity. 
Each heatmap visualizes the conditional preference preservation ratio, measuring how consistently users maintain their dominant preference label when moving from a base category ($\boldsymbol{y}$-axis) to a compared category ($\boldsymbol{x}$-axis). }
\label{fig:app_motiv}
\vspace{-0.4cm}
\end{figure*}

\vspace{1mm}
\noindent\textbf{Intra-domain preference heterogeneity analysis.}
\textbf{Figure~\ref{fig:app_motiv}} further extends our analysis of intra-domain preference heterogeneity across additional domains and preference criteria.
Specifically, it reports price-based preservation in the Home domain (Figure~\ref{fig:app_motiv}-(a)), and rating-based preservation in the Cloth domain (Figure~\ref{fig:app_motiv}-(b)).
The heatmaps visualizes the conditional preference preservation ratio, which measures how consistently users retain their dominant preference labels when transitioning between base and compared categories within the same domain.

\input{Table/table-d_desc-prompt.tex}
\input{Table/table-app-opensourceLLM.tex}
\input{Table/table-app-noLLM.tex}

\FloatBarrier

\clearpage
\input{Table/table-eq1_all}

%% file: Table/table-persona_db.tex

\begin{table}[ht]
\caption{Example user persona DB constructed from metadata}
\centering
\footnotesize 
\renewcommand{\arraystretch}{1.5}
\begin{tabularx}{\columnwidth}{X}
\hline
\rowcolor[HTML]{F2F2F2} 
\textbf{User persona database (e.g., Home)} \\ \hline
[ \newline 
\{ "User ID": "AEXXX", \newline
"price\_affiliated\_group": \{ \newline
\quad "Kitchen \& Dining": "L", "Vacuums \& Floor Care": "L", \newline
\quad "Storage \& Organization": "H", "Home Décor Products": "M", \newline
"rating\_score\_preferred\_group": \{ \newline
\quad "Kitchen \& Dining": "M", "Home Décor Products": "H", \newline
\quad "Bedding": "M", "Vacuums \& Floor Care": "H", \newline
\quad "Storage \& Organization": "H" \}, \newline
"rating\_nums\_preferred\_group": \{ \newline
\quad "Kitchen \& Dining": "L", "Home Décor Products": "H", \newline
\quad "Bedding": "L", "Vacuums \& Floor Care": "L", "Storage \& Organization": "M"\}, \newline
"cats\_familiarity": \{ \newline
\quad "Bedding": "L", "Home Décor Products": "H", \newline
\quad "Kitchen \& Dining": "L",  \newline
\quad "Storage \& Organization": "L", "Vacuums \& Floor Care": "L" \}, \newline
"cats\_interaction\_diversity": "L" \}, \newline
\{ "User ID": "AGXXX", \newline
"price\_affiliated\_group": \{ \newline
\quad "Kitchen \& Dining": "H", "Vacuums \& Floor Care": "L", \newline
\quad "Furniture": "H", "Bedding": "L", "Storage \& Organization": "H" \}, \newline
\quad ... \}  \newline
] \\ \hline
\end{tabularx}
\label{tab:user_persona_db}
\vspace{-3mm}
\end{table}

%% file: Table/table-d_desc-prompt.tex
\begin{table}[t]
\caption{System prompt for domain description generation}
\label{tab:domain_description_prompt}
\centering
\scriptsize
\renewcommand{\arraystretch}{1.4} 
\begin{tabularx}{\columnwidth}{l X} 
\hline
\rowcolor[HTML]{F2F2F2} 
\textbf{Component} & \textbf{Content} \\ \hline
\textbf{Task Description} & You are an e-commerce platform analyst that generates a concise and general domain description using only the provided item metadata samples. The generated description is used to condition subsequent LLM-based user persona construction. \\ \hline

\textbf{Input Format} & 
\texttt{\{ "Domain": "<Domain Name>",} \newline
\texttt{\ \ "sampled\_item\_list": \{ "item\_id": "<title || category || description>" \} \}} \\ \hline

\textbf{Generation Rules} & 
Write a **Domain Description** that summarizes the overall characteristics of the domain in 2–3 sentences. \\ \hline

\textbf{Output Format} & 
\texttt{\{ "Domain": "<Domain Name>",} \newline
\texttt{\ \ "Domain Profile": \{ "Domain Description": "..." \} \}} \\ \hline

\textbf{Example Result} & 
\texttt{\{ "Domain": "Electronics",} \newline
\texttt{\ \ "Domain Profile": \{ "Domain Description": "The Electronics domain encompasses a wide range of gadgets and devices designed for personal, professional, and entertainment purposes. It includes innovative technologies like memory cards, wireless headphones, portable projectors, computing accessories, power solutions, and GPS devices. This domain focuses on functionality, connectivity, durability, and modernized design to meet the needs of technologically savvy consumers."\} \}} \\ \hline
\end{tabularx}
\end{table}

%% file: Table/table-app-opensourceLLM.tex
\begin{table}[t]
\centering
\caption{Performance of \ours\ with different LLMs}
\label{tab:app-openLLM}
\vspace{-0.25cm}
\resizebox{1\columnwidth}{!}{ 
\begin{tabular}{cc|c|cc} 
\toprule
\textbf{Domain} 
& \textbf{Metrics (@5)} 
& \textbf{PPA}
& \textbf{Multi-TAP (Qwen-14B)}
& \textbf{Multi-TAP (GPT-4o)} \\ 
\midrule

\multirow{2}{*}{Home $\rightarrow$ Toys} 
& HR   
& 1.35 $\pm$ 0.05 
& \underline{1.46 $\pm$ 0.03} 
& \textbf{1.56 $\pm$ 0.24} \\
& NDCG 
& 0.36 $\pm$ 0.02 
& \underline{0.39 $\pm$ 0.02} 
& \textbf{0.47 $\pm$ 0.03} \\ 
\midrule

\multirow{2}{*}{Toys $\rightarrow$ Home} 
& HR   
& 2.98 $\pm$ 0.04 
& \underline{3.06 $\pm$ 0.22} 
& \textbf{3.11 $\pm$ 0.15} \\
& NDCG 
& 0.80 $\pm$ 0.01 
& \textbf{0.89 $\pm$ 0.08} 
& \underline{0.88 $\pm$ 0.03} \\ 
\bottomrule
\end{tabular}
}
\vspace{-0.3cm}
\end{table}

%% file: Table/table-app-noLLM.tex
\begin{table}[t]
\centering
\caption{Effect of LLM-generated personas}
\label{tab:app-noLLM}
\vspace{-0.25cm}
\resizebox{0.9\columnwidth}{!}{ 
\begin{tabular}{cc|cc} 
\toprule
\textbf{Domain} 
& \textbf{Metrics (@5)} 
& \textbf{ w/o LLM personas}
& \textbf{Multi-TAP} \\ 
\midrule

\multirow{2}{*}{Home $\rightarrow$ Toys} 
& HR   & \underline{1.33 $\pm$ 0.23} & \textbf{1.56 $\pm$ 0.24} \\
& NDCG & \underline{0.36 $\pm$ 0.03} & \textbf{0.47 $\pm$ 0.03} \\ 
\midrule

\multirow{2}{*}{Toys $\rightarrow$ Home} 
& HR   & \underline{1.87 $\pm$ 0.14} & \textbf{3.11 $\pm$ 0.15} \\
& NDCG & \underline{0.51 $\pm$ 0.06} & \textbf{0.88 $\pm$ 0.03} \\ 
\bottomrule
\end{tabular}
}
\vspace{-0.1cm}
\end{table}

%% file: Table/table-eq1_all.tex
\begin{table*}[t]
\centering
\caption{Performance comparison (\%) of \ours\ on all remaining cross-domain pairs from the Amazon dataset}
\label{table:app_amazon_results}
\vspace{-0.2cm}
\resizebox{0.82\textwidth}{!}{
\begin{tabular}{c|c|ccc|ccccc|c} 
\toprule
\textbf{Domain} 
& \multirow{2}{*}{\textbf{\begin{tabular}[c]{@{}c@{}}Metrics\\(@5)\end{tabular}}}
& \multicolumn{3}{c|}{\textbf{Single-Domain Methods}} 
& \multicolumn{5}{c|}{\textbf{Cross-Domain Methods}} 
& \multirow{2}{*}{\textbf{\ours}} \\ 
\textbf{(Source $\to$ Target)} 
& 
& BPRMF & NGCF & LightGCN
& UniCDR & CD-IMF & CUT- LightGCN & PPA & DGCDR
&  \\ \midrule

\multirow{2}{*}{\begin{tabular}[c]{@{}c@{}}Cloth $\to$ Toys\end{tabular}} 
& HR    & 1.09 & 0.91 & 0.91 & 1.09 & 1.32 & 1.61 & 1.52 & \underline{1.92} & \textbf{2.33} \\
& NDCG  & 0.30 & 0.16 & 0.35 & 0.31 & 0.38 & 0.46 & 0.52 & \underline{0.55} & \textbf{0.76} \\
\hline

\multirow{2}{*}{\begin{tabular}[c]{@{}c@{}}Toys $\to$ Cloth\end{tabular}} 
& HR    & 2.18 & 0.91 & 2.18 & 2.96 & 3.05 & 3.28 & \underline{3.46} & 3.33 & \textbf{4.01} \\
& NDCG  & 0.58 & 0.22 & 0.57 & 0.75 & 0.75 & 0.79 & \underline{0.81} & 0.80 & \textbf{1.11} \\
\midrule

\multirow{2}{*}{\begin{tabular}[c]{@{}c@{}}Elec $\to$ Cloth\end{tabular}} 
& HR    & 1.52 & 0.50 & 1.18 & 1.44 & 1.34 & 1.49 & \underline{1.58} & 1.52 & \textbf{2.18} \\
& NDCG  & 0.44 & 0.16 & 0.38 & 0.39 & 0.37 & 0.48 & \underline{0.49} & 0.47 & \textbf{0.60} \\
\hline

\multirow{2}{*}{\begin{tabular}[c]{@{}c@{}}Cloth $\to$ Elec\end{tabular}} 
& HR    & 0.86 & 0.68 & 1.03 & 0.95 & 1.10 & 1.28 & \textbf{2.58} & 2.31 & \textbf{2.58} \\
& NDCG  & 0.25 & 0.23 & 0.29 & 0.26 & 0.31 & 0.36 & \underline{0.60} & 0.50 & \textbf{0.61} \\
\midrule

\multirow{2}{*}{\begin{tabular}[c]{@{}c@{}}Elec $\to$ Sports\end{tabular}} 
& HR    & 0.50 & 1.01 & 1.69 & 1.55 & 1.78 & 1.81 & \textbf{2.03} & 1.84 & \uline{2.01} \\
& NDCG  & 0.31 & 0.31 & 0.68 & 0.42 & 0.48 & 0.54 & \textbf{0.71} & 0.57 & \uline{0.65} \\
\hline

\multirow{2}{*}{\begin{tabular}[c]{@{}c@{}}Sports $\to$ Elec\end{tabular}} 
& HR    & 1.72 & 1.37 & 0.86 & 1.30 & 1.48 & 1.66 & \underline{2.40} & 2.11 & \textbf{2.88} \\
& NDCG  & 0.58 & 0.47 & 0.21 & 0.36 & 0.41 & 0.47 & \underline{0.63} & 0.52 & \textbf{0.90} \\
\midrule

\multirow{2}{*}{\begin{tabular}[c]{@{}c@{}}Elec $\to$ Toys\end{tabular}} 
& HR    & 0.47 & 0.71 & 1.19 & 1.10 & 1.35 & 1.72 & \underline{3.34} & 2.71 & \textbf{3.98} \\
& NDCG  & 0.27 & 0.31 & 0.45 & 0.30 & 0.38 & 0.50 & \underline{1.12} & 0.91 & \textbf{1.34} \\
\hline

\multirow{2}{*}{\begin{tabular}[c]{@{}c@{}}Toys $\to$ Elec\end{tabular}} 
& HR    & 1.21 & 0.72 & 1.70 & 1.45 & 1.62 & 1.88 & \underline{4.17} & 3.01 & \textbf{4.33} \\
& NDCG  & 0.50 & 0.25 & 0.49 & 0.40 & 0.46 & 0.54 & \underline{1.30} & 0.89 & \textbf{1.41} \\
\midrule

\multirow{2}{*}{\begin{tabular}[c]{@{}c@{}}Home $\to$ Cloth\end{tabular}} 
& HR    & 0.67 & 1.01 & 1.52 & 1.60 & 1.78 & 1.95 & \underline{3.39} & 2.65 & \textbf{3.71} \\
& NDCG  & 0.30 & 0.41 & 0.51 & 0.44 & 0.49 & 0.55 & \underline{0.91} & 0.79 & \textbf{0.98} \\
\hline

\multirow{2}{*}{\begin{tabular}[c]{@{}c@{}}Cloth $\to$ Home\end{tabular}} 
& HR    & 1.02 & 0.68 & 1.36 & 1.25 & 1.42 & 1.60 & \textbf{3.41} & 2.71 & \uline{3.24} \\
& NDCG  & 0.30 & 0.14 & 0.32 & 0.34 & 0.40 & 0.46 & \textbf{0.82} & 0.62 & \uline{0.72} \\
\midrule

\multirow{2}{*}{\begin{tabular}[c]{@{}c@{}}Home $\to$ Sports\end{tabular}} 
& HR    & 1.36 & 0.84 & 1.35 & 1.40 & 1.55 & 1.75 & \underline{2.71} & 2.12 & \textbf{2.74} \\
& NDCG  & 0.33 & 0.26 & 0.58 & 0.38 & 0.43 & 0.51 & \underline{0.87} & 0.71 & \textbf{0.90} \\
\hline

\multirow{2}{*}{\begin{tabular}[c]{@{}c@{}}Sports $\to$ Home\end{tabular}} 
& HR    & 1.01 & 0.68 & 1.19 & 1.20 & 1.35 & 1.55 & \underline{2.56} & 2.24 & \textbf{3.41} \\
& NDCG  & 0.29 & 0.17 & 0.32 & 0.33 & 0.38 & 0.45 & \underline{0.65} & 0.51 & \textbf{0.78} \\
\midrule

\multirow{2}{*}{\begin{tabular}[c]{@{}c@{}}Sports $\to$ Toys\end{tabular}} 
& HR    & 0.91 & 0.72 & 0.91 & 0.95 & 1.10 & 1.32 & \underline{2.00} & 1.70 & \textbf{2.18} \\
& NDCG  & 0.24 & 0.15 & 0.28 & 0.26 & 0.31 & 0.38 & \underline{0.61} & 0.51 & \textbf{0.75} \\
\hline

\multirow{2}{*}{\begin{tabular}[c]{@{}c@{}}Toys $\to$ Sports\end{tabular}} 
& HR    & 1.27 & 1.09 & 2.21 & 2.10 & 2.35 & 2.75 & \textbf{3.46} & 3.01 & \textbf{3.46} \\
& NDCG  & 0.46 & 0.28 & 0.77 & 0.58 & 0.65 & 0.78 & \textbf{0.99} & 0.83 & \uline{0.95} \\
\bottomrule
\end{tabular}
}
\vspace{-0.3cm}
\end{table*}